\documentstyle[aps,graphicx]{revtex}
\begin{document}

\preprint{\vbox{\hbox{UCD-97-24}\hbox{DOE-ER40757-107}
                \hbox{UTEXAS-HEP-97-21}
                \hbox{OITS-639}
                \hbox{OSURN-331}
                  }}
\draft
\title {Multilepton signatures of gauge mediated \\SUSY breaking at LEPII}
\author{ Kingman Cheung$^{1,2,3}$, Duane A. Dicus$^{1,2} $,
\\ B. Dutta$^{4} $, and S. Nandi$^{5}$ }
\address{
$^{1}$ Center for Particle Physics, University of Texas, Austin, TX 78712\\
$^{2}$ Department of Physics, University of Texas, Austin, TX 78712\\
$^{3}$ Department of Physics, University of California at Davis, Davis, 
CA 95616\\
$^{4}$ Institute of  Theoretical Science, University of Oregon, Eugene, OR
97403\\
$^{5} $ Department of Physics, Oklahoma State University, Stillwater, OK 74078}
\date{October, 1997}
\maketitle

\begin{abstract} 
In the framework of gauge-mediated supersymmetry 
breaking models pair production of the lightest neutralinos, scalar leptons,
or charginos at LEPII gives rise to interesting signals involving
multilepton final states and missing energy.
In the parameter space where the scalar tau, $\tilde{\tau}_1$, is the 
next-to-lightest supersymmetric particle, we identify three interesting 
regions, which give rise to distinctly different final states:
(i) 2 $\tau$-leptons plus missing energy, (ii) 4 charged leptons plus
missing energy where in some regions
all four are $\tau$-leptons, or (iii) six charged leptons, of which four are 
$\tau$-leptons and the other two are electrons or muons, plus missing energy.  
We study in detail the size of these regions in the parameter space of 
gauge-mediated models and give cross section contours in these regions
for various LEPII energies.
We also discuss the possibility of chargino-pair production at LEPII.
\end{abstract}


\newpage
\section{Introduction} 
Supersymmetry is usually assumed to be broken in a hidden sector and
then communicated to the observable sector, but how this communication
happens is an area of active research.  Most commonly, this
communication is via gravity, the result of which is that the soft
terms are introduced at around the Planck scale ($\sim 10^{18}$
GeV). This mode of communication, however, has the problem of
uncontrolled flavor violation, which needs to be harnessed by some
symmetry.  Further, if these terms are introduced at the Planck scale
or at the string scale, one needs a definite unification group from
the GUT scale ($\sim 2\times10^{16}$ GeV) up to the Planck scale,
because in the minimal supersymmetric standard Model (MSSM) the
couplings unify at the GUT scale.  Consequently, the radiative
corrections to the fields depend on the particular choice of the
unifying group. Recently, another class of models has become very
popular. In these models communication takes place via gauge
interactions which have the standard model (SM) gauge symmetry
\cite{DN} or a new gauge symmetry \cite{DMN}.  
%
%
The soft terms are introduced at
a scale around 100 TeV.  Comparing wth the supergravity-motivated
models, gauge-mediated models have fewer parameters because the
coefficients of the trilinear terms in the potential ($A$ terms) are
zero at the boundary. We will concentrate on this type of models in
this paper.

The gravitino is always the LSP in these gauge-mediated supersymmetry breaking
(GMSB) models.  For the NLSP spot there is
 a competition between the lighter scalar tau (stau) 
$\tilde{\tau}_1$ 
and the lightest neutralino $\tilde{\chi}^0_1$. The winner
is decided by the input parameters.  Signals from the production of SUSY
particles can distinguish which is the NLSP. The phenomenology when the 
lightest neutralino is the NLSP and the resulting signals of hard photons 
plus missing energy (due to the gravitinos) have been discussed extensively
for the Tevatron and for LEPII 
\cite{DWR,swy,bkw,akkmm,dwt,bbct,BPM,rtm,BR,MO,aab,KT,sb}.
If the scalar tau $\tilde{\tau}_1$ is the NLSP its decay to a $\tau$-lepton
and a gravitino gives rise to a signature of extra $\tau$-leptons 
plus missing energy, without any hard photons. 
The production of neutralinos, scalar electrons (selectrons), scalar muons
(smuons), and staus as well as 
their signatures have been discussed in the literature by us 
and others \cite{{we1},{B},{akm},{we2},{dn}}. 
In this paper, we elaborate on the processes specific to 
LEPII where the stau is the NLSP
and determine the regions in the parameter space that can be probed by
LEPII.

For a beam energy $E_{\rm beam}=E_{CM}/2$ at LEPII 
the parameter space can be divided into three distinct regions:

\begin{itemize}

\item Region I: $E_{\rm beam} >m_{\tilde{\chi}^0_1} > 
 m_{\tilde{e}_1}>m_{\tilde\tau_1}$, where the lightest selectron $\tilde{e}_1$
is usually the right-handed one $\tilde{e}_R$.  
In this region, the final state will be $4e$, $4\mu$,
$4\tau$, $2e2\mu$, $2e2\tau$, or $2\mu2\tau$ plus missing energy.

\item Region II: $E_{\rm beam} >m_{\tilde{e}_1}> m_{\tilde{\chi}^0_1}
 > m_{\tilde\tau_1}$.  Here the signals  are
$4\tau$ or 6 charged leptons plus missing energy.

\item Region III: $m_{\tilde{e}_1} >E_{\rm beam} > m_{\tilde{\chi}^0_1}>
 m_{\tilde\tau_1}$.  In this region, the
signal will be $4\tau$ plus missing energy.

\end{itemize}

In the above classification we have neglected
the three-body decay mode of the selectron. This mode
appears when the selectron mass is less than the
neutralino mass (region I) and the final state
has 6 leptons plus missing energy. Hence, if the
three-body decay mode of selectron dominates over
the two-body decay mode, the
boundary between the region I and region II
becomes less transparent. In section IV we show
graphically the regions where the three-body decay mode
dominates over the two body mode.
Here the smuon mass is assumed to be equal to the selectron mass.  
In each of these regions a pair of staus can be directly produced 
but it gives a signal of only two $\tau$-leptons plus missing energy, 
which not only has a smaller cross section but also suffers from $WW$ 
background.

Our objective is to make a detailed study of the GMSB parameter
space to see how probable these regions are and to investigate the
corresponding signals. 
The organization of the paper is as follows.
In section II we discuss the GMSB parameter space and how we calculate the 
sparticle mass spectrum and how to identify the regions in which 
$\tilde\tau_1$ is the NLSP.  In section III we discuss the production processes
contributing to the signals and the final states in the regions I, II, and III.
In section IV, we give our results and describe in detail the signals in
regions I, II, and III.
Section V contains a brief discussion on the prospect
of chargino-pair production at LEPII.
Section VI contains our conclusions. 
In an appendix we point out that the cross section for
$e^+ e^- \to \tilde{\chi}^0_1 \tilde{\chi}^0_1$ is more sensitive to
a polarized electron beam in GMSB models than in supergravity models.
Thus a polarized electron beam might be employed to distinguish between
these two supersymmetry models.  

\section {GMSB parameter space with $\tilde\tau_1$ as NLSP}

In  GMSB models the sparticle masses depend on five parameters:
$M,\Lambda, n, \tan\beta$, and sign$(\mu)$.  $M$
is the messenger scale, $M=\lambda \langle s \rangle$, 
where $\langle s \rangle$ is the VEV of the scalar
component of superfield in the hidden sector and $\lambda$ is the Yukawa
coupling. The parameter $\Lambda$ is equal to $\langle F_s \rangle/
\langle s \rangle$, where $ \langle F_s \rangle$ is the VEV 
of the auxiliary component of the superfield. $F_s$ can be of order
of the intrinsic SUSY breaking scale $F$. In GMSB models,
$\Lambda$ is taken to be around 100 TeV so that the colored superpartners have
masses around 1 TeV or less. The parameter $n$ is fixed by the choice of the
messenger sector. The messenger-sector representations should be vector-like
(for example, 
$5+{\bar 5}$ of
$SU(5)$, $10+{\bar {10}}$ of $SU(5)$ or $16+{\bar{16}}$ of $SO(10)$) 
so that their masses are well above the electroweak scale. 
They are also chosen to transform as a GUT multiplet in order
not to affect the gauge coupling unification in MSSM. These facts  
restrict $n(5+{\bar
5})\le 4$, or $n(10+{\bar {10}})\le 1$ in
$SU(5)$,  and $n(16+{\bar {16}})\le 1$ in an
$SO(10)$ GUT for the messenger sector (one
${10}+{\bar{10}}$ pair corresponds to
$n(5+{\bar 5})$=3). The parameter $\tan\beta$ is the usual ratio of the up
($H_u$) and down ($H_d$) type Higgs VEVs. The parameter
$\mu $ is the coefficient in the bilinear term, $\mu H_uH_d$, in the
superpotential while another parameter $B$ is defined to be the coefficient in
the bilinear term, $B\mu H_uH_d$, in the potential. In general, 
$\mu$ and $B$ depend on the details
of the SUSY breaking in the hidden sector. We demand that the electroweak
symmetry is broken radiatively, 
which determines $\mu^2$ and $B$ in terms of 
other parameters of the theory. Thus we are left with five independent
parameters,
$M,\Lambda,n,\tan\beta$ and sign($\mu$). The soft SUSY breaking gaugino and the
scalar masses at the messenger scale $M$ are given by \cite{SPM}
\begin{center}
$\tilde M_i(M) = n\,g\left({\Lambda\over M}\right)\,
{\alpha_i(M)\over4\pi}\,\Lambda$
\end{center} and
\begin{center}
$\tilde m^2(M) = 2 \,n\, f\left({\Lambda\over M}\right)\,
\sum_{i=1}^3\, k_i \, C_i\,
\biggl({\alpha_i(M)\over 4\pi}\biggr)^2\,
\Lambda^2 $
\end{center} 
where $\alpha_i\; (i=1-3)$ are the three SM gauge couplings and
$k_i=1,1,3/5$ for SU(3), SU(2), and U(1), respectively. The $C_i$ are zero for
gauge singlets, and 4/3, 3/4, and $(Y/2)^2$ for the fundamental representations
of
$SU(3)$ and $SU(2)$ and $U(1)_Y$ respectively 
(with $Y$ defined by $Q=I_3+Y/2)$. Here $n$ corresponds to
$n(5+{\bar 5})$. $g(x)$ and $f(x)$ are messenger scale threshold functions 
with $x=\Lambda/M$ \cite{SPM}.

We calculate the SUSY mass spectrum using the appropriate RGE equations
\cite{BBO} with the boundary conditions given by the equations above and
vary the five free parameters. 
 For the messenger sector, we choose  $5+{\bar 5}$ of SU(5), and varied
$n(5+{\bar 5})$ from 1 to 4. In addition to the current experimental bounds on
the superpartner masses, the rate for $b\rightarrow s\gamma $ 
restricts $\mu<0$ \cite{{dwt},{bbct},{ddo}}. 
In the absence of late
inflation, cosmological constraints put an upper bound on the gravitino mass of
about $10^4$ eV \cite{pp}, which restricts $M/\Lambda=1.1-10^4$. It
is  found that for $n=1$ and $\tan\beta \alt 25$  the lightest neutralino
$\tilde{\chi}^0_1$ is the NLSP \cite{dwt,BPM}.  As $\tan\beta$ increases 
further,  $\tilde\tau_1$ becomes the NLSP.
For $n\ge2$, $\tilde\tau_1$ is the NLSP even for low values of
$\tan\beta$ ($\tan\beta \agt 2$), and for $n\ge3 $
$\tilde\tau_1 $ is naturally the NLSP for most of the parameter space.  

\section{Collider signals with $\tilde\tau_1$ as the NLSP} 

In this section, we
discuss the various processes that could give rise to observable signals at
LEPII.  We  divide the parameter space into the three   
regions based on the mass hierarchies of the superpartners.  The final
states can be different in these three regions.  By searching for the signals 
of regions I, II, and III  the LEPII experiments should 
find SUSY or be able to exclude parts
or all of these regions in the parameter space. 

In region I ($E_{\rm beam} >m_{\tilde{\chi}^0_1} > 
 m_{\tilde{e}_1}>m_{\tilde\tau_1}$), the following production processes
are kinematically allowed
\begin{eqnarray} 
e^{+}e^{-} &\to& \tilde{\chi}^0_1 \tilde{\chi}^0_1 \;, \label{p1}\\ 
e^{+}e^{-} &\to& \tilde{e}^+_1 \tilde{e}^-_1 \;, \;\; 
   \tilde{\mu}^+_1\tilde{\mu}^-_1 \;,  \label{p2} \\ 
e^{+}e^{-} &\to& \tilde{\tau}^+_1 \tilde{\tau}^-_1  \label{p3}\;.
\end{eqnarray}
The neutralino can decay to any of the sleptons:
\begin{eqnarray}
\tilde{\chi}^0_1&\rightarrow&e\tilde e_1\rightarrow  ee\tilde G \;,
 \label{4}\\
\tilde{\chi}^0_1&\rightarrow&\mu\tilde \mu_1\rightarrow  \mu\mu\tilde G \;,
  \label{5} \\
\tilde{\chi}^0_1&\rightarrow&\tau\tilde \tau_1\rightarrow  \tau\tau\tilde G \;.
\end{eqnarray} 
Thus, the interesting final states in region I are $4e$, $4\mu$,
$4\tau$, $2e2\mu$, $2e2\tau$ or $2\mu2\tau$ plus missing energy from the pair
production of neutralinos. 

In this region the selectron and the smuon also have a three-body
decay mode, the rate of which is comparable to the two-body decay mode just
discussed \cite{akm2}.  The selectron (smuon) decays into an electron
(muon) and an off-shell neutralino, which further decays into a pair of
$\tau$-leptons and missing energy.  The final states are
$2e+4\tau$ and $2\mu+4\tau$ plus missing energy.

In region II ($E_{\rm beam} >m_{\tilde{e}_1}> m_{\tilde{\chi}^0_1}
 > m_{\tilde\tau_1}$), all the processes (\ref{p1})--(\ref{p3}) are allowed.
There are, however, additional final states because the selectron and smuon 
can decay into the neutralino first:
\begin{eqnarray}
\tilde e_1&\rightarrow& e\tilde{\chi}^0_1 \rightarrow 
  e\tau\tilde\tau_1\rightarrow e\tau\tau\tilde G  \;, \\
\tilde \mu_1&\rightarrow& \mu \tilde{\chi}^0_1
  \rightarrow \mu\tau\tilde\tau_1\rightarrow \mu\tau\tau\tilde G \;.
\end{eqnarray}  
Given the choices the selectron or smuon will more likely decay through the
$\tilde{\chi}^0_1$ because it involves the weak interaction 
whereas the direct decay to a lepton and a gravitino is gravitational. 
For the same reason the $\tilde{\chi}^0_1$ will decay to lepton-slepton
rather than $\gamma$ + $\tilde G$.
Thus, the interesting final states are $4\tau$ plus missing energy from 
process (\ref{p1}) or 6 charged leptons, four of which are taus,
plus missing energy from process (\ref{p2}).

In region III ( $m_{\tilde{e}_1} >E_{\rm beam} > m_{\tilde{\chi}^0_1}>
 m_{\tilde\tau_1}$), only the processes (\ref{p1}) and (\ref{p3}) are 
kinematically allowed.  The only allowed decays are
\begin{eqnarray}
\tilde{\chi}^0_1&\rightarrow&\tau\tilde \tau_1\rightarrow\tau\tau\tilde G \;,\\
\tilde \tau_1&\rightarrow&\tau\tilde G \;,
\end{eqnarray}
and so the final states are $4\tau$ or $2\tau$ with missing energy.

The cross-sections for the processes (\ref{p1})--(\ref{p3})
are different.  Usually the neutralino pair production cross-section is
the largest of the three processes in GMSB models because the 
$\tilde{e}_R$ exchanged in the $t$-channel is often light.  
Selectron pair production 
involves the $s$-channel $\gamma$ and $Z$ exchanges and the 
$t$-channel neutralino 
exchange while smuon and stau pair production involves only the $s$-channel
$\gamma$ and $Z$ exchanges. So the production rate of a 6 lepton final state
with 2 muons and 4 taus is in general different from that with 2 electrons
and 4 taus.   For an illustration of the relative magnitudes
we show all these pair production  cross-sections in  Table 1 for
four different scenarios.

In order that the signatures considered above be seen we assume that 
the decay length of the NLSP, $\tilde{\tau}_1$, into the $\tau$-lepton 
and the gravitino is less than the size of the detector. 
If the decay length is significantly larger than the dimension
of the detector, the $\tilde{\tau}_1$ will travel through the detector and be
seen as  a heavy charged particle. So, for example, rather than seeing
four $\tau$-leptons with missing energy, the signal would be two $\tau$-leptons
and two heavy charged particles without any missing energy.
The decay width $\Gamma$ for $\tilde{\tau}_1$ to
$\tau \tilde{G}$ is
\begin{equation}
\label{**}
\Gamma = \frac{M_{\tilde{\tau}_1}^5}{16 \pi F^2} \;
\end{equation} 
where $\sqrt{F}$ is the SUSY breaking scale. The probability that a
particle can travel a distance $x$ before it decays is
$P(x)=1 - \exp^{-x/L}$, where $L$ is the decay length given by \cite{BPM,akm}
\begin{equation} 
L = (2 \times 10^{-16} {\rm m} )\; \left( \frac{1\;{\rm GeV}}{\Gamma}
\right ) \;
\sqrt{\frac{E_{\tilde{\tau}_1}^2}{m_{\tilde{\tau}_1}^2} -1 } \;.
\end{equation} 
The last factor is the Lorentz factor.  At LEPII, the
energy $E_{\tilde{\tau}_1}$ of the scalar tau is only $1-2$ times its mass, 
and so the Lorentz factor is only of order 1.  
To set a criterion for the $\tilde{\tau}_1$ to decay within the detector, 
we require $L \alt 2 \;{\rm m}$, which gives a condition on $\Gamma$:
\begin{equation}
\Gamma \agt 10^{-16} \;{\rm GeV} \;.
\end{equation} 
Assuming $M_{\tilde{\tau}_1} \simeq 50$ GeV, the condition
$\Gamma \agt 10^{-16}$ GeV implies
\begin{equation}
\sqrt{F} \alt 5 \times 10^{5} \; {\rm GeV} \;.
\end{equation}
In summary, if $\sqrt{F} \agt 5 \times 10^{5}$ GeV the decay of the
$\tilde{\tau}_1$ is likely to be outside the detector; otherwise if
$\sqrt{F}\alt 5 \times 10^{5}$ GeV the decay is within the
detector. 
If $F_s\sim F$, then, this leads to the condition
$\Lambda\agt \sqrt{\lambda\over{M/\Lambda}}$ 500 TeV for the decay
to take place outside the detector, where $\lambda$ is the coupling by which
the superfield $S$ 
is coupled to the messenger fields. Similarly for smaller masses of
selectrons and smuons the decays into $e\tilde G$ and $\mu\tilde G$ 
(Eq. (\ref{4}) and Eq. (\ref{5}))
 can also happen outside the detector. In all
these situation there would be two track of heavy charged particles instead of 
high $p_T$ leptons in the final state.

\section {GMSB parameter space for the regions I, II and II}  

In this section, we describe in detail our investigation of the 
GMSB parameter space that gives regions I, II, and III and of how
probable these regions are.  As mentioned before, the GMSB parameters are
$M, \Lambda, n, \tan\beta$ and
sign$(\mu)$. We take sign$(\mu)$ to be negative in order 
to satisfy the $b\rightarrow s\gamma$ constraint and vary $n$ from 1 to 4, 
$\tan\beta$ from 1 to 50, and $\Lambda$ from 10 TeV to 100 TeV. 
The value of $M/\Lambda$ is chosen to be 1.1 or 100, as representative values.
Larger values of $M/\Lambda$ give results similar to those of $M/\Lambda=100$.
We also impose the experimental constraints
$m_h>60$ GeV (where $h$ is the lighter neutral scalar Higgs boson), 
$m_{\tilde g}>200$ GeV (where $\tilde g $ is the gluino), and
$ m_{\tilde\tau_1}>45$ GeV. We summarize our results 
in the $\Lambda$-$\tan\beta$ plane for several values of $n$ and
$M/\Lambda$. Figures 1(a)--1(d) are for $\sqrt s=$172 GeV, Fig. 2(a)--2(c)
for $\sqrt s=$183 GeV, and Fig. 3(a)--3(c) for $\sqrt s=$194 GeV.
In Fig. 1(a), we have used $n=1$ and $M=1.1\Lambda$. 
The curve $r\equiv m_{\tilde{\tau}_1}/m_{\tilde{\chi}^0_1}=1$ corresponds 
to where the $m_{\tilde{\tau}_1}=m_{\tilde{\chi}^0_1}$.
The region above this curve is for $\tilde{\tau}_1$ as the NLSP and 
we are primarily interested in this region. 
The darkest shaded region is excluded by the lighter neutral scalar Higgs boson
mass $m_h <60$ GeV.  We also draw the contour of $m_{\tilde{g}}=200$ GeV, to 
the left of which is also excluded.
However, the most severe constraint comes from the demand that the 
stau mass to be greater than 45 GeV.   The region on the right of the curve 
satisfies the constraint, as indicated by the arrow. The solid line
labelled ``neutralino=$E/2$'' represents the contour along which 
$m_{\tilde{\chi}^0_1}$ is equal to the beam energy 
(which in this case is $E_{\rm beam}=E/2=86$ GeV). 
The allowed region for $\tilde{\chi}^0_1$-pair production is on the left of 
this contour.  
Another solid line labelled ``selectron=$E/2$'' corresponds to the contour 
along which $m_{\tilde{e}_R}=E/2$. This line separates region II from 
region III.  
The available regions at this energy are labelled II and III, as shown. 
There is no region I in this case.
We see that there is a considerable range of $\Lambda\simeq 32 - 57$ TeV and 
$\tan\beta \simeq 20-37$ for which regions II and III are allowed.
When the ratio $M/\Lambda$ increases,  the selectron and stau masses 
increase because of the increase in the threshold function $f(\Lambda/M)$;
while the neutralino mass decreases due to the decrease in the threshold 
function  $g(\Lambda/M)$. 
Consequently, region III would enlarge while region II would shrink.
On the other hand, as $m_{\tilde{\chi}^0_1}$ decreases in order 
for $m_{\tilde\tau_1}$ to remain the NLSP, larger values of $\tan\beta$ is
needed.  Thus the net result for the allowed region is an increase in the 
range of $\Lambda$ but a decrease in the range of $\tan\beta$. 
In Fig. 1(a), we also show the cross-section contours for 
neutralino-pair production by the dashed lines.
In regions II and III, the cross section varies from 0.05 pb to
0.65 pb.  As $\Lambda$ decreases  the cross section first increases to a
maximum and then decreases. 
This observation can be understood in terms of the two masses
$m_{\tilde{\chi}^0_1},\; m_{\tilde{e}_R}$ and the neutralino-mixing matrix
element $N_{11}$.  
The cross section is roughly proportional to $|N_{11}|^4$. 
The smaller the $\Lambda$, the smaller the masses 
$m_{\tilde{\chi}^0_1}$ and $m_{\tilde{e}_R}$  will be, which give more phase
space to the process $e^+e^- \to \tilde{\chi}^0_1 \tilde{\chi}^0_1$, and so
the cross section will increase.  However, as $\Lambda$ decreases the nature
of the lightest neutralino changes from more gaugino-like to more
higgsino-like, which causes a decrease in $N_{11}$.  
Consequently, the combined effect makes the cross section increase to 
a maximum when $\Lambda$ decreases from a very large value.  
After this maximum the cross section decreases as $\Lambda$ decreases 
further in the allowed region where the dominant contribution comes
from the $\tilde{e}_R$ $t$-exchange diagram, and the cross section
depends mainly on $N_{11}$. In the excluded region (very small $\Lambda$)
the Higgsino component becomes large,  the contribution from
the $Z$-exchange diagram is dominant, and the cross section is large.
But this region is excluded by the experimental constraint on the Higgs mass.

Next we consider Fig. 1(b). Here we show the effect of changing $n$ to 2 but
keeping $M/\Lambda$ and $\sqrt s$ the same as in Fig. 1(a). 
The available regions in this case are II and I but not region III.
The reason for this change is that the increase in $n$ causes the 
neutralino mass $m_{\tilde{\chi}^0_1}$ to increase more than the
selectron mass $m_{\tilde{e}_R}$. 
The cross sections are somewhat smaller ($\sim 0.1$ to 0.35 pb) in the 
available regions. 
When the ratio $M/\Lambda$ is increased to 100, region I disappears but 
region III 
appears again for the reason mentioned above, as shown in Fig. 1(c). 
Note that the cross sections are larger here because in this case of
$M/\Lambda=100$, the matrix element $|N_{11}|$ decreases much less rapidly
than that in the case of $M/\Lambda=1.1$ when $\Lambda$ decreases.
It therefore implies that the cross section can increase to a higher maximum
in this case of $M/\Lambda=100$ when $\Lambda$ decreases.
In Fig. 1(d), we show the effect of increasing $n$ to 4. It is clear that
the available regions appear at smaller $\Lambda$ and lower $\tan\beta$.
We show regions I, II, and III and the cross sections for $E_{CM}=\sqrt s= 
183$ GeV in Fig. 2 and those for $\sqrt{s}=194$ GeV in Fig. 3.

We have shown in Figs. 1--3 the allowed regions I, II, and III
in the $\Lambda$-$\tan\beta $ plane and cross sections for 
$e^+e^-\rightarrow \tilde{\chi}^0_1 \tilde{\chi}^0_1$ within these regions.
Failure to find the unique signals of this cross-section to some level would
further restrict the $\Lambda$-$\tan\beta$ parameters. But what do these
restrictions mean in terms of more physical parameters such as the masses?
Figures 1--3 show that $m_{\tilde{\chi}^0_1}$ and $m_{\tilde{e}_R}$ 
are roughly independent of $\tan\beta$ while $m_{\tilde\tau_1}$ is 
definitely not. For any one of
choices of $n$ and $M/\Lambda$ a plot of $m_{\tilde e_R}$ as a function of
$m_{\tilde{\chi}^0_1}$ is a unique line where $m_{\tilde\tau_1}$ depends on 
$\tan\beta$, as shown in Fig. 4(a).
The intersection of the selectron line with the diagonal
separates region I ($m_{\chi^0_1} > m_{\tilde e_R}$) from region II
($m_{\chi^0_1} < m_{\tilde e_R}$) while the intersection of stau with the
diagonal bounds the lower end of region II since we always require
$m_{\tilde\tau_1} < m_{\chi^0_1}$. 
The $\tan\beta=$ 20 region is further bounded
by the requirement that $m_{\tilde\tau_1}$ be larger than 45 GeV. These
boundaries can then be applied to the cross-section as in Fig. 4(b). 
For example, for the $n$ and $M/\Lambda$ values used here, an experimental 
determination that
the cross-section is less than 0.25 pb for $\sqrt{s}=183$ GeV would rule
out region II entirely and restrict $m_{\tilde{\chi}^0_1}$ to be larger 
than 75 GeV as
can be seen from Fig. 4(b) or smaller than 66, 55, or 51 GeV for $\tan\beta$
equal 10, 15, or 20 as can be seen from Fig. 4(a). 
(In Fig. 4(b), the three cross
section lines for each beam energy show the slight 
dependence of the cross section on
$\tan\beta$.)

Figures 5(a) and 5(b) are similar to Figs. 4(a) and 4(b) but for $n=2$ and
$M/\Lambda=100$ where region III is allowed but region I is not. The separation
between regions II and III is fixed by the ratio of $m_{\tilde{e}_R}$ 
to the beam energy; the beam energy is taken as 97 GeV for the separation 
shown. The limit on the left side of region II is given by $m_{\tilde\tau_1} 
<  m_{\chi^0_1}$ for 
$\tan\beta=$16 but by $m_{\tilde\tau_1}>$ 45 GeV for $\tan\beta=$20 or 25. As
before an upper bound on the cross-section can put limits on
$m_{\tilde{\chi}^0_1}$. For example at $\sqrt s=$194 GeV, an upper limit on the
cross-section for the final states of region II of 0.65 pb implies 
$m_{\tilde{\chi}^0_1} >$ 64 GeV  or
$m_{\tilde{\chi}^0_1} <$ 54 GeV from Fig. 5(b) and from Fig. 5(a)
$m_{\tilde{e}_R} >$ 80 GeV or less than 72 GeV if
$\tan\beta$=20, but gives no constraint if $\tan\beta=16$ or 25.

{}From the above results we conclude that there is a considerable region of the
GMSB parameter space for which regions I, II, and III are available at LEPII
energies. The production of the superparticles ($\tilde{\chi}^0_1$, 
$\tilde e_R$,  $\tilde\mu_R$, 
$\tilde\tau_1$) in these available regions will give rise to the final states
with two, four, or six charged leptons plus missing energy as discussed in
section III. If such final states are not observed at LEPII, then the upper
limit on the cross-section times the branching ratios for these final states
will further reduce the allowed regions, and will set new limits on the
allowed masses for $\tilde{\chi}^0_1$, 
$\tilde e_R$, $\tilde\mu_R$ and $\tilde\tau_1$.

Now we discuss the three-body decay mode of the selectron and smuon.
As mentioned in Sec. II, in region I there is also a possibility of
three-body decays of the selectron and smuon, $\tilde{e} \to e \tau
\tilde{\tau}$ and $\tilde{\mu} \to \mu \tau \tilde{\tau}$, through a
virtual neutralino, with the subsequent decay $\tilde{\tau} \to \tau
\tilde{G}$, which might be larger than the two-body decays, $\tilde{e}
\to e \tilde{G}$ and $\tilde{\mu} \to \mu \tilde{G}$.  We investigate
two of our cases where Region I exists: (i) $n=2$ and $M=1.1\Lambda$,
and (ii) $n=4$ and $M=100\Lambda$.  For $n=4$ and $M=100\Lambda$,
$\sqrt {F_s}(\equiv \sqrt {{{\Lambda M}\over\lambda}})$ is between
$160\over{\sqrt \lambda}$ and $200\over{\sqrt \lambda}$ TeV for Region
I as can be seen in Fig. 1(d), 2(c), and 3(c).  Taking $F$ to be equal
to $F_s$ and $\sqrt\lambda=1$, we have calculated the rates for the
above two-body and three-body decays.  Figures 6(a) and 6(b) represent
the contour in the $\Lambda-\tan\beta$ plane where these two rates are
equal. From Fig. 6(a) corresponding to $n=2$ and $M=1.1\Lambda$ , we
see that the three-body decay rate is larger than two-body decay rate
for $\tan\beta \agt 7$; whereas for $n=4$ and $M=100\Lambda$ the
corresponding values are $\tan\beta \agt 4$.  Note that the experimental 
limit on the mass of the lighter stau rules out (i) for 
$\tan\beta \agt 25$ and (ii) for $\tan\beta \agt 20$.

\section {chargino pair production at LEPII} 

In the GMSB parameter space where $\tilde{\chi}^0_1$ is the NLSP,
the chargino is always heavier than 100 GeV and, therefore,
 cannot be pair produced at LEPII \cite{akkmm}.
On contrary, in the parameter space where the $\tilde\tau_1$ as the NLSP, 
there are regions in parameter space where the chargino is light
enough to be pair produced at LEPII. 
In Figs. 7(a) and 7(b), we show the region in the $\Lambda$-$\tan\beta$ plane 
where the chargino mass is less than the beam energy 
for $n=2$ and $4$ with the experimental constraints as before. 
For $n=2$ and $M/\Lambda=1.1$, we get a  triangular region around 
$\Lambda\simeq 20$ TeV and $\tan\beta \simeq 13 - 22$ as shown in Fig. 7(a).
For n=4 the region expands
somewhat around $\Lambda\simeq 13$ TeV and $\tan\beta \simeq 5 - 20$.  
Thus, there are regions in parameter space where not only 
the chargino mass is light enough to be pair produced at LEPII but also 
the other conditions are satisfied. The mass hierarchy in this region is 
\begin{equation} 
m_{\tilde{\chi}^0_{2}} \geq m_{\tilde{\chi}^\pm_1} > m_{\tilde{e}_1,\tilde
{\mu}_1} >m_{\tilde{\chi}^0_1} >m_{\tilde \tau_1} \;.
\end{equation} 
The scalars $\tilde e_1$ and $\tilde \mu_1$ are essentially the right-handed 
selectron and smuon and have equal masses. The only allowed decay
mode for the chargino is: 
$\tilde{\chi}^+_1 \rightarrow \nu_\tau\tilde\tau_1$: see table 2. 
(Since the lighter $\tilde e_1$ and $\tilde \mu_1$ 
are essentially right handed the branching ratios of the other two
kinematically allowed decay modes, $\nu_e\tilde e_1$ and $\nu_\mu\tilde \mu_1$,
are essentially zero.)    The
$\tilde\tau_1$ then decays to a $\tau$-lepton and a gravitino with a 100\% 
branching ratio. The final states arising from the decays of the produced 
chargino pair
are two oppositely charged $\tau$ leptons, accompanied by large ${\rlap/E}_T$.
The corresponding background from the $W$-pair production is $\sigma.B\sim
$0.18 pb, 0.21 pb,  and 0.23 pb 
at the CM energies $\sqrt s=$172 GeV, 183 GeV,
and 194 GeV, respectively. In Table 2 we show the cross sections for 
chargino piar production at LEPII for some allowable scenarios. 
We see that the signal is considerably larger than the background.

\section{Conclusion} 

In this work, we have studied in detail the 
parameter space in GMSB models where the $\tilde\tau_1$ is the NLSP.  We have
identified three regions in the $\Lambda$-$\tan\beta$ plane based on the 
mass hierarchies of the supersymmetric particles: $\tilde{\chi}^0_1$,
$\tilde{e}_R$, and $\tilde{\tau}_1$.
Different mass hierarchies produce different final
states: $2\tau$-leptons plus missing energy, 4 charged leptons plus missing 
energy, or six charged leptons plus missing energy. 
None of these signals involve hard photons.
We have also calculated the contours of the cross sections for pair
production of the lightest neutralinos in these regions for various
LEPII energies.
For practical purposes we have shown the same cross sections as a function
of the neutralino mass.
The cross section is in general of order 0.5 pb, which may be large enough
to be observed at LEPII.  
The LEPII experiments should be able to find these SUSY signals or 
be able to rule out part or all of these regions in the parameter space.
Figures 1--5 summarize our findings.

In addition, we pointed out that there is a small region of parameter space 
where charginos may be light enough to be pair produced at LEPII. 
It gives clean signature of $\tau$-leptons  with large missing energy.

\section*{Acknowledgement}
This work was supported in part by the  US Department of Energy Grants No.
DE-FG013-93ER40757, DE-FG02-94ER40852, DE-FG03-96ER-40969, and 
DE-FG03-91ER40674.

\newpage

\begin{center} {\bf Appendix }
\end{center} 

There is an interesting polarization effect present in GMSB models.  In
these models, because the particles get their masses via gauge interactions,
the masses of $\tilde e_L$ and $\tilde e_R$ are naturally split with
$\tilde e_R$ being much lighter than $\tilde e_L$.
In supergravity models these masses are probably not split so much
since that would require a large value of $A$ or $\mu$.
The cross section for
$e^{+}e^{-} \rightarrow\tilde{\chi}^0_1 \tilde{\chi}^0_1$
is dominated by selectron exchange in the $t$-channel.  The coupling of
the electron and lightest neutralino to the right-handed selectron,
in both theories, is twice the coupling of the electron and lightest
neutralino to the left-handed selectron, times a factor that depends on
the neutralino mixing.  A polarized electron beam could be used to
compare the cross section from $\tilde e_L$ exchange to that from
$\tilde e_R$ \cite{pol}.  Neglecting neutralino mixing the ratio would
be one-sixteenth in supergravity but much smaller in GMSB.
Neutralino mixing is small at large $\Lambda$ but cannot be
neglected at smaller $\Lambda$ and its effect is to suppress further
the cross section from $\tilde e_L$.  For example the further suppression
is more than a factor of ten for the scenarios of Table 2.

At present LEPII does not provide polarized beams.
But the beams may be polarized
at future electron or muon colliders and the difference
in the cross sections might be used to determine the relative selectron
masses and thus distinguish between GMSB and supergravity.

A similar effect might be seen in pair production of selectrons at high
energy where the $t$-channel $\tilde{\chi}^0_1$ exchange will dominate.
Of course if the selectrons are being produced their masses can probably
be determined by more direct means.

\newpage
\begin{center} {\bf TABLE CAPTIONS}\end{center}
\begin{itemize}
\item[Table 1~:]{Cross sections for $\tilde{\chi}^0_1 \tilde{\chi}^0_1$, 
$\tilde{\tau}_1^+  \tilde{\tau}_1^-$, $\tilde{e}_R^+ \tilde{e}_R^-$, and
$\tilde{\mu}_R^+ \tilde{\mu}_R^-$ pair production at LEPII and the masses
for the lightest neutralino $\tilde{\chi}^0_1$, lightest stau $\tilde{\tau}_1$,
and the lightest selectron $\tilde{e}_R$ in
four different scenarios. The smuon and the selectron masses are the same. }

\item[Table 2~:] { Mass spectrum for the superpartners and 
$\tilde{\chi}^+_1 \tilde{\chi}^-_1$ pair 
production cross-sections at the LEPII for $\sqrt{s}=172,\; 183$, and 
194 GeV.}
\end{itemize}

\begin{center} {\bf FIGURE CAPTIONS}\end{center} 
\begin{itemize}

\item[Fig. 1:] 

Regions I, II, and III and cross section contours (dashed lines) for 
$e^+ e^- \to \tilde{\chi}^0_1 \tilde{\chi}^0_1$ at center-of-mass energy
$E_{CM} = 172$ GeV in the $\Lambda$-$\tan\beta$ plane of the gauge-mediated
models with experimental constraints from the lighter neutral Higgs boson,
scalar tau, and gluino.  The regions are defined by the mass hierarchy of the
sparticle masses as follows: 
I: $E_{\rm beam} > m_{\tilde{\chi}^0_1} >
    m_{\tilde{e}_R} > m_{\tilde{\tau}_1}$, 
II:  $E_{\rm beam} > m_{\tilde{e}_R} > m_{\tilde{\chi}^0_1} > 
     m_{\tilde{\tau}_1}$, and 
III:  $m_{\tilde{e}_R} > E_{\rm beam} > 
   m_{\tilde{\chi}^0_1} >  m_{\tilde{\tau}_1}$.
Above the contour $r\equiv m_{\tilde{\tau}_1}/m_{\tilde{\chi}^0_1}=1$ the
$\tilde{\tau}_1$ is the NLSP.
The imposed constraints are: (i) the lighter neutral Higgs boson mass
$m_h > 60$ GeV, the darkest shaded region is excluded, (ii) the gluino
mass $m_{\tilde{g}}>200$ GeV, the region on the right of the contour
$m_{\tilde{g}}=200$ GeV is allowed, and (iii) the stau mass $m_{\tilde{\tau}_1}
>45$ GeV, the allowed region is indicated by an arrow.
In the figure, $E=E_{CM}$, (a) $n=1, M=1.1\Lambda$,
  (b) $n=2, M=1.1\Lambda$,
  (c) $n=2, M=100\Lambda$, and 
  (d) $n=4, M=100\Lambda$

\item[Fig. 2:]
Same as Fig. 1, except for $E_{CM}=183$ GeV and we only show
(a) $n=1, M=1.1\Lambda$,
(b) $n=2, M=100\Lambda$, and
(c) $n=4, M=100\Lambda$.

\item[Fig. 3:]
Same as Fig.1, except for $E_{CM}=194$ GeV and we only show
(a) $n=1, M=1.1\Lambda$,
(b) $n=2, M=100\Lambda$, and
(c) $n=4, M=100\Lambda$.

\item[Fig. 4:]
(a)The lighter selectron mass $m_{\tilde{e}_R}$ and the lighter stau mass
$m_{\tilde{\tau}_1}$ versus the lightest neutralino mass 
$m_{\tilde{\chi}^0_1}$ for $n=2, \; M=1.1\Lambda$ at $E_{CM}=194$ GeV.  The 
selectron mass $m_{\tilde{e}_R}$ curve is roughly independent of $\tan\beta$
while the stau mass $m_{\tilde{\tau}_1}$ curves are shown for $\tan\beta=
10,\,15,\,20$.  
The upper boundary of region I is given by $m_{\tilde{\chi}^0_1} < E_{\rm beam}
=E_{CM}/2$.  The separation between region I and II is determined by 
$m_{\tilde{e}_R} \stackrel{<}{\scriptstyle >}  m_{\tilde{\chi}^0_1}$.
The  lower boundary of region II is given by $m_{\tilde{\chi}^0_1} > 
m_{\tilde{\tau}_1}$ or $m_{\tilde{\tau}_1}>45$ GeV, whichever gives a larger
$m_{\tilde{\chi}^0_1}$, which is about 51, 55, and 66 GeV for 
$\tan\beta=20,15,10$, respectively.
Care has to be taken for smaller $E_{CM}$ because the upper boundary of
region I will shift towards smaller $m_{\tilde{\chi}^0_1}$.
The diagonal straight line is $m_{\tilde{\tau}_1, \tilde{e}_R}=
m_{\tilde{\chi}^0_1}$. 

(b)
Cross sections in pb for $e^+ e^- \to \tilde{\chi}^0_1 \tilde{\chi}^0_1$
at $E_{CM}=172, 183$, and 194 GeV versus $m_{\tilde{\chi}^0_1}$ for 
$n=2$ and $M=1.1\Lambda$.  The boundaries of regions I and II are the same
as in part (a).  The cross section has mild dependence on $\tan\beta$
and for each center-of-mass energy we show the cross section for 
$\tan\beta =10$ (dot-dashed), $15$ (solid), and $20$ (dashed).  

\item[Fig. 5:]
Same as Fig. 4, but with $n=2$, $M=100\Lambda$, and regions II and III
appears but not region I.  The upper boundary of region III is given by
$m_{\tilde{\chi}^0_1} < E_{\rm beam}$.  The separation between regions II and
III are determined by 
$m_{\tilde{e}_R} \stackrel{<}{\scriptstyle >}  E_{\rm beam}$.
The lower boundary of region II is given by $m_{\tilde{\chi}^0_1} > 
m_{\tilde{\tau}_1}$ or $m_{\tilde{\tau}_1}>45$ GeV, whichever gives a larger
$m_{\tilde{\chi}^0_1}$, which is about 54, 64, and 71 GeV for 
$\tan\beta=20,16,25$, respectively.
Care has to be taken for smaller $E_{CM}$ because the upper boundary of
region III and the separation between II and III  will both shift towards 
smaller $m_{\tilde{\chi}^0_1}$. 
The diagonal straight line is $m_{\tilde{\tau}_1, \tilde{e}_R}=
m_{\tilde{\chi}^0_1}$. 

\item[Fig. 6:] Figure showing the contour along which the two-body and 
three-body decay rates are equal for (a) $n=2,\; M=1.1 \Lambda$ and 
(b) $n=4,\; M=100\Lambda$. The region above the contour is dominated by 
the three-body decay.

\item[Fig. 7:]
Figure showing the allowed region for chargino-pair production at LEPII with
$\sqrt{s}=183$ GeV for 
(a) $n=2,\; M=1.1 \Lambda$ and (b) $n=4,\; M=1.1\Lambda$.
The constraints shown are: $m_{\tilde{\tau}_1}>45$ GeV, 
$r\equiv m_{\tilde{\tau}_1}/m_{\tilde{\chi}^0_1}<1$, and 
$m_{\tilde{\chi}^+_1} < E/2$.  Here $E=E_{CM}$.
\end{itemize}

\newpage
\begin{center}  Table 1 \end{center}
\begin{center}\begin{tabular}{|c|c|c|c|c|}  \hline  &Scenario 1&Scenario
2&Scenario3&Scenario 4\\\hline  &$\Lambda=16$ TeV,&$\Lambda=13$TeV,
&$\Lambda=25$
TeV,&$\Lambda=28$ TeV,\\ &n=3,$M= 40\Lambda$&n=4, $M=10\Lambda$&n=2,
$M=20\Lambda$&n=2,
$M=40\Lambda$\\
&$\tan\beta$=16&$\tan\beta$=15&$\tan\beta$=20&$\tan\beta$=18\\\hline 
$M_{\tilde{\chi}^0_1}$(GeV)&56&57&63&72\\\hline 
$m_{\tilde\tau_1}$(GeV)&50&51&53&65\\\hline
$m_{\tilde e_R}$(GeV)&67&64&77&85\\\hline
$\sqrt s$=172 GeV&&&&\\
$\sigma_{e^+e^-\rightarrow \tilde{\chi}^0_1 \tilde{\chi}^0_1}$(pb)
 &0.65&0.43&0.59&0.33\\
$\sigma_{e^+e^-\rightarrow {\tilde\tau _1}^+{\tilde\tau
_1}^-}$&0.475&0.465&0.434&0.252\\
$\sigma_{e^+e^-\rightarrow\tilde e_R^+\tilde
e_R^-}$&0.246&0.165&0.103&0.003\\
$\sigma_{e^+e^-\rightarrow\tilde\mu_R^+\tilde\mu_R^-}$&0.234&0.291&0.085
&0.003\\\hline
$\sqrt s$=183 GeV&&&&\\
$\sigma_{e^+e^-\rightarrow \tilde{\chi}^0_1 \tilde{\chi}^0_1}$(pb)
 &0.68&0.45&0.65&0.43\\
$\sigma_{e^+e^-\rightarrow {\tilde\tau _1}^+{\tilde\tau
_1}^-}$&0.456&0.448&0.423&0.275\\
$\sigma_{e^+e^-\rightarrow\tilde e_R^+\tilde e_R^-}$&0.326&0.212&0.191&0.052\\
$\sigma_{e^+e^-\rightarrow\tilde\mu_R^+\tilde\mu_R^-}$&0.263&0.309&0.132
&0.042\\\hline
$\sqrt s$=194 GeV&&&&\\
$\sigma_{e^+e^-\rightarrow \tilde{\chi}^0_1 \tilde{\chi}^0_1}$(pb)
  &0.70&0.46&0.70&0.52\\
$\sigma_{e^+e^-\rightarrow {\tilde\tau _1}^+{\tilde\tau
_1}^-}$&0.433&0.426&0.406&0.285\\
$\sigma_{e^+e^-\rightarrow\tilde e_R^+\tilde
e_R^-}$&0.396&0.254&0.277&0.118\\
$\sigma_{e^+e^-\rightarrow\tilde\mu_R^+\tilde\mu_R^-}$&0.277&0.316&0.166
  &0.083\\\hline
\end{tabular}
\end{center}
\newpage
\begin{center}  Table 2 \end{center}
\begin{center}
\begin{tabular}{|c|c|c|c|c|}  \hline  &Scenario 1&Scenario 2&Scenario3&Scenario
4\\\hline  &$\Lambda=13.5$ TeV,&$\Lambda=17$TeV,&$\Lambda=20$
TeV,&$\Lambda=22$ TeV,\\ &n=4,$M= 1.1\Lambda$&n=3, $M=1.1\Lambda$&n=2,
$M=80\Lambda$&n=2,
$M=3\Lambda$\\
&$\tan\beta$=16&$\tan\beta$=17&$\tan\beta$=18&$\tan\beta$=20\\\hline  m$_h
$(GeV)&111&111&107&110\\\hline m$_{H^{\pm}}$&153&160&199&183\\\hline 
m$_A$&130&115&182&165\\\hline m$_{\tilde{\chi}^0_1}$&59&59&48&51\\\hline 
m$_{\tilde{\chi}^0_2}$&112&107&84&88\\
 \hline m$_{\tilde{\chi}^0_3}$&119&129&182&161\\\hline 
m$_{\tilde{\chi}^0_4}$&225&215&212&203\\\hline
m$_{\tilde{\chi}^{\pm}}$&82,225&85,216&80,214&82,204\\\hline 
m$_{\tilde{\tau}_{1,2}}$&57,120&56,123&46,128&50,132\\\hline 
m$_{\tilde{e}_{1,2}}$&64,117&66,118&68,119&69,124\\\hline
m$_{\tilde{\nu}}$&84&88&88&94\\\hline  m$_{\tilde {\rm
t}_{1,2}}$&404,458&405,457&326,401&386,439\\\hline  m$_{\tilde {\rm
b}_{1,2}}$&403,416&403,418&334,359&384,405\\\hline  m$_{\tilde {\rm
u}_{1,2}}$&407,416&408,418&348,360&395,406\\\hline  m$_{\tilde {\rm
d}_{1,2}}$&408,423&410,426&350,369&395,413\\\hline  m$_{\tilde
g}$&609&554&356&397\\\hline
$\mu$&-105&-114&-167&-146\\\hline
$\sigma_{e^+e^-\rightarrow \tilde{\chi}^+_1 \tilde{\chi}^-_1}$(pb)&&&&\\
$\sqrt s=$172 GeV&0.85&0.38&0.41&0.40\\\hline
$\sqrt s=$183 GeV&1.02&0.68&0.50&0.50\\\hline
$\sqrt s=$194 GeV&1.04&0.76&0.57&0.54\\\hline
\end{tabular}
\end{center}

\begin{figure}[th]
\begin{center}
\includegraphics[width=3.2in]{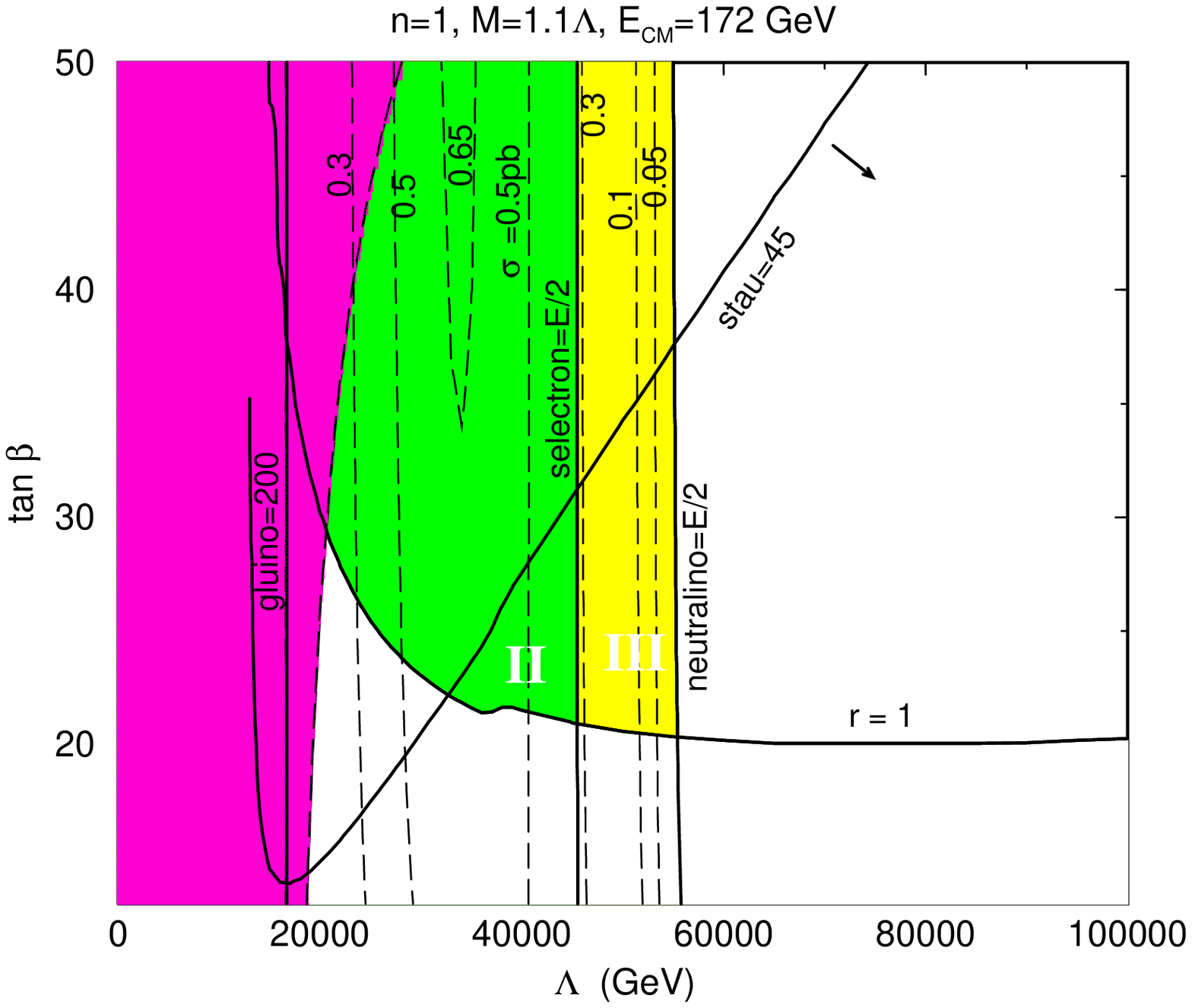}
\includegraphics[width=3.2in]{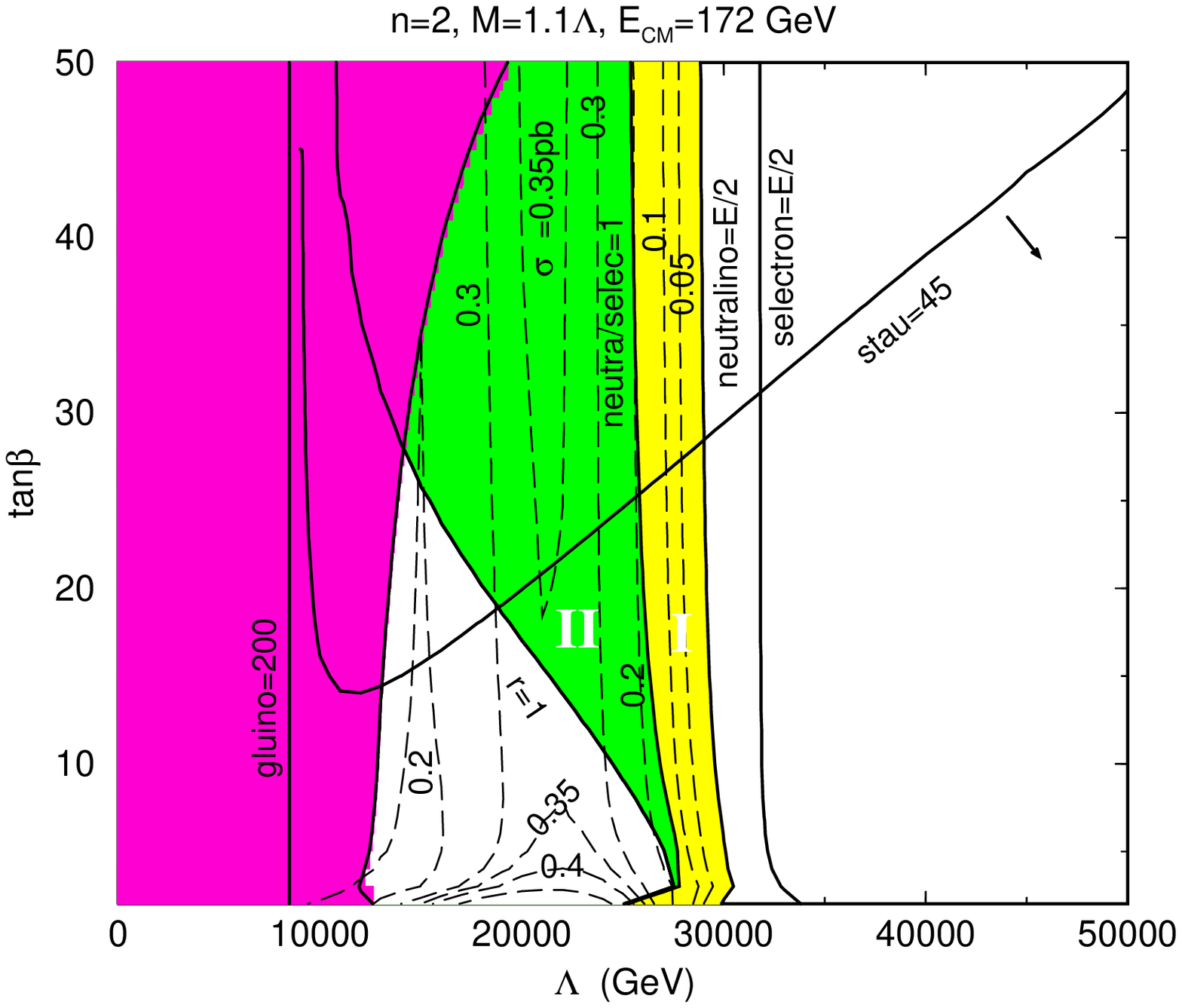}
\includegraphics[width=3.2in]{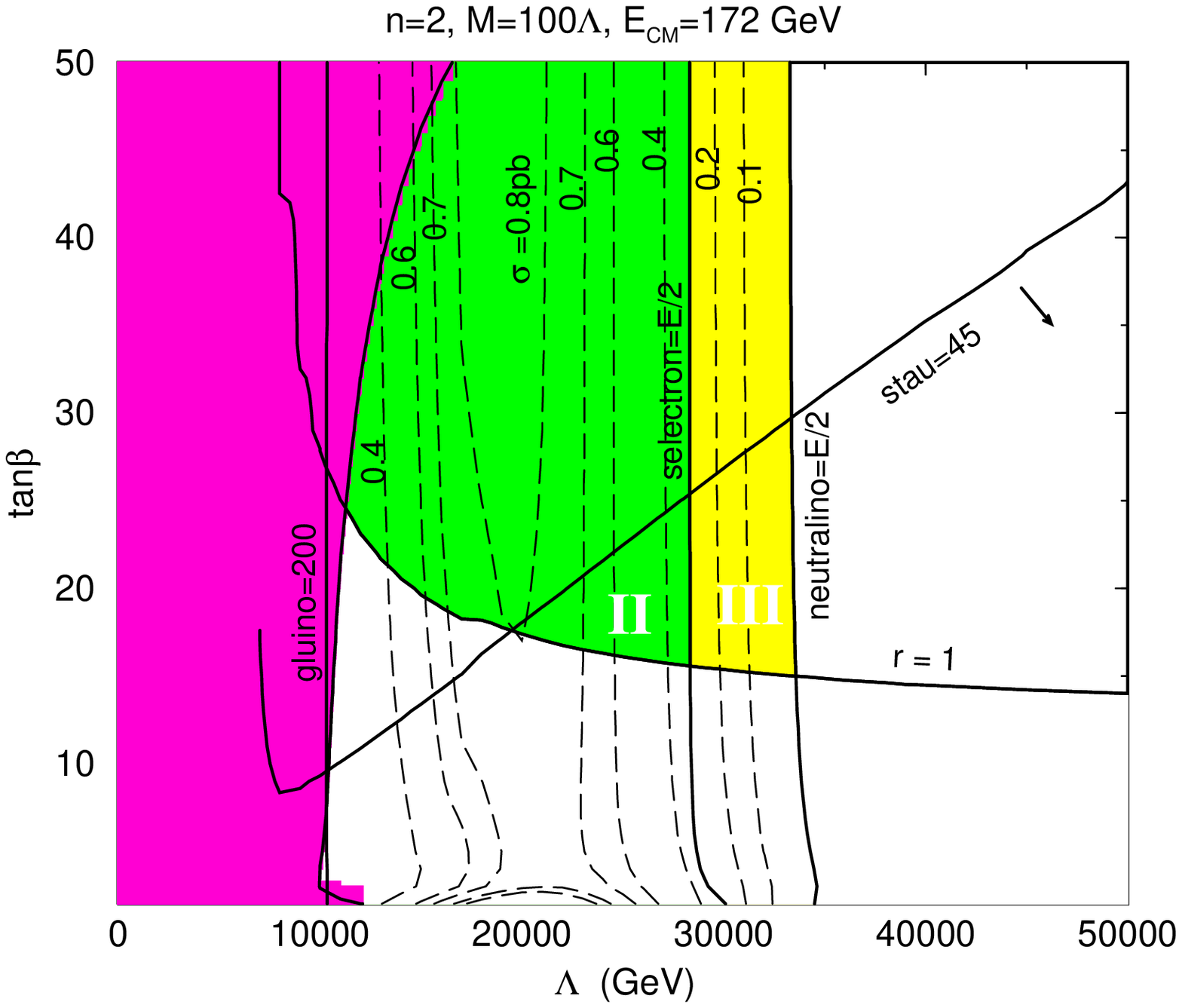}
\includegraphics[width=3.2in]{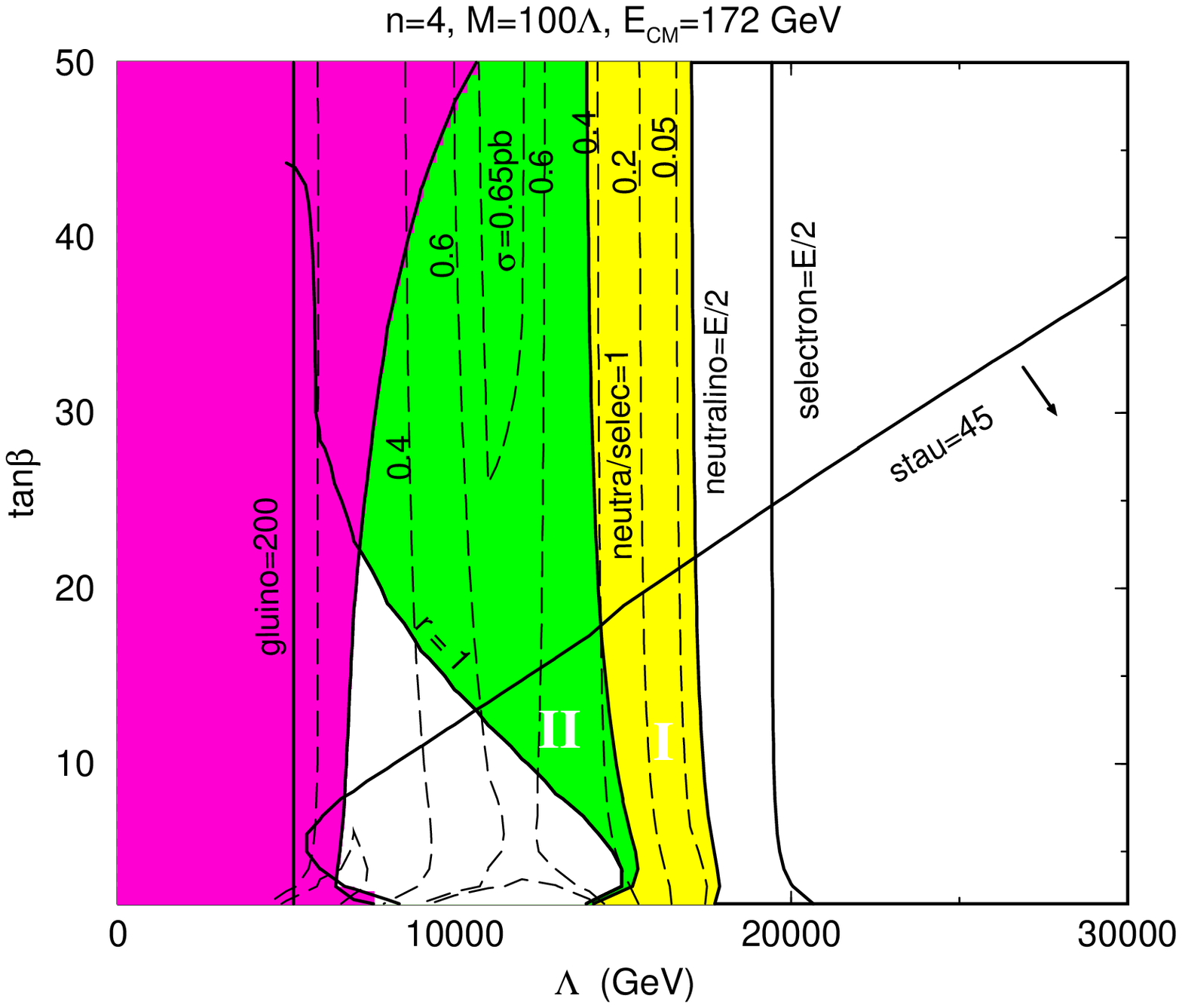}
\end{center}
\caption{}
\end{figure}

\begin{figure}[th]
\leavevmode
\begin{center}
\includegraphics[height=2.85in]{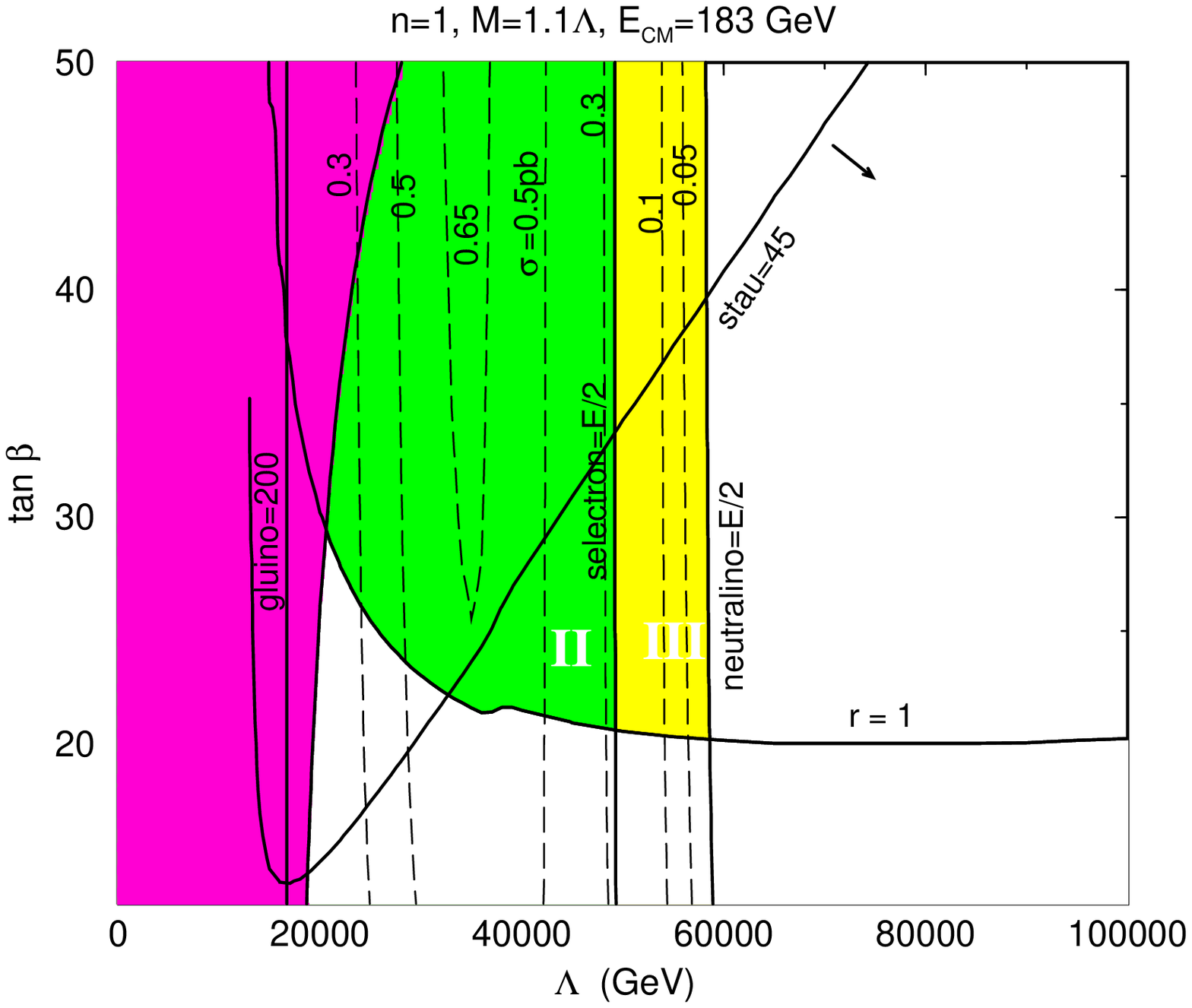}
\includegraphics[height=2.85in]{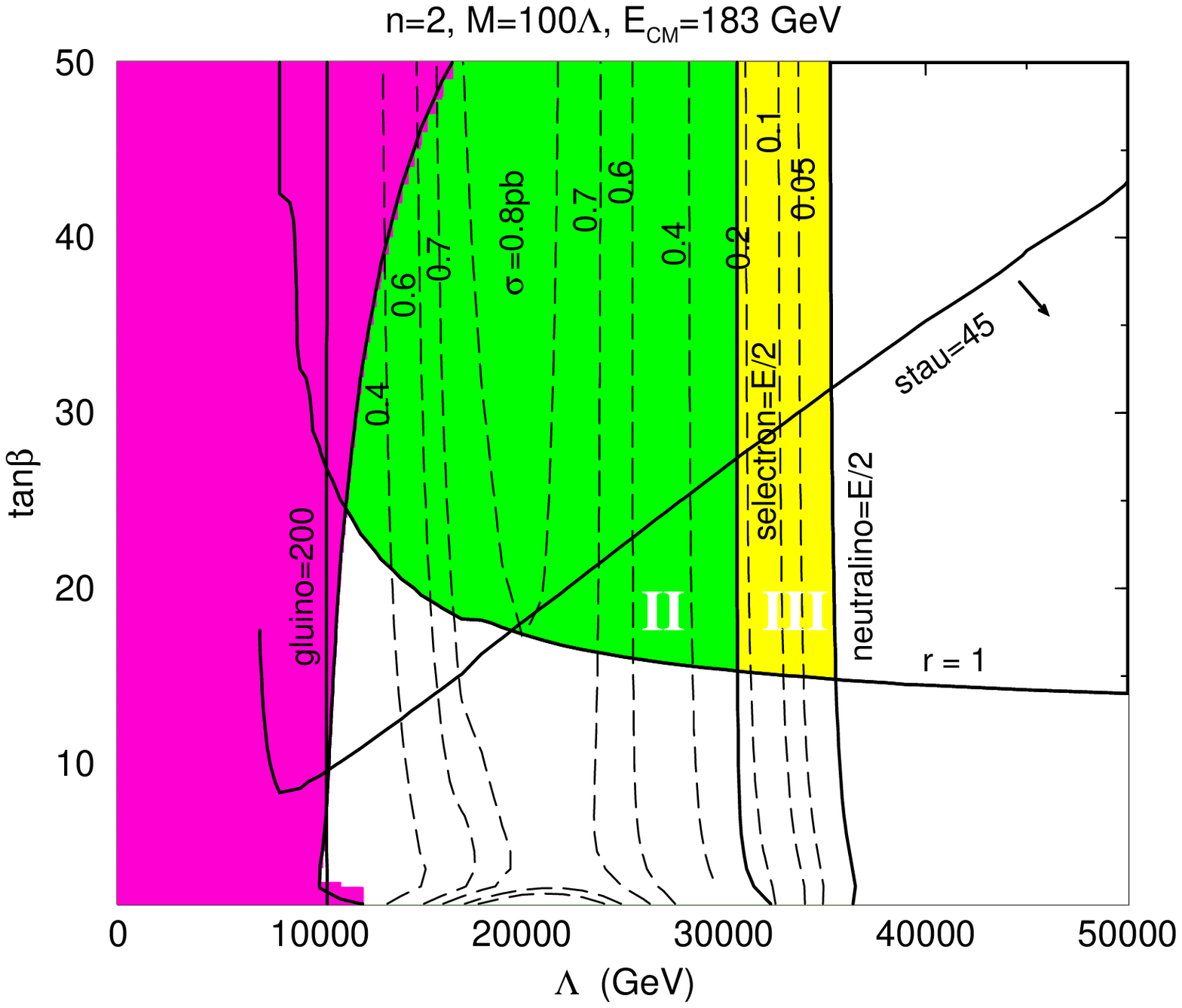}
\includegraphics[height=2.85in]{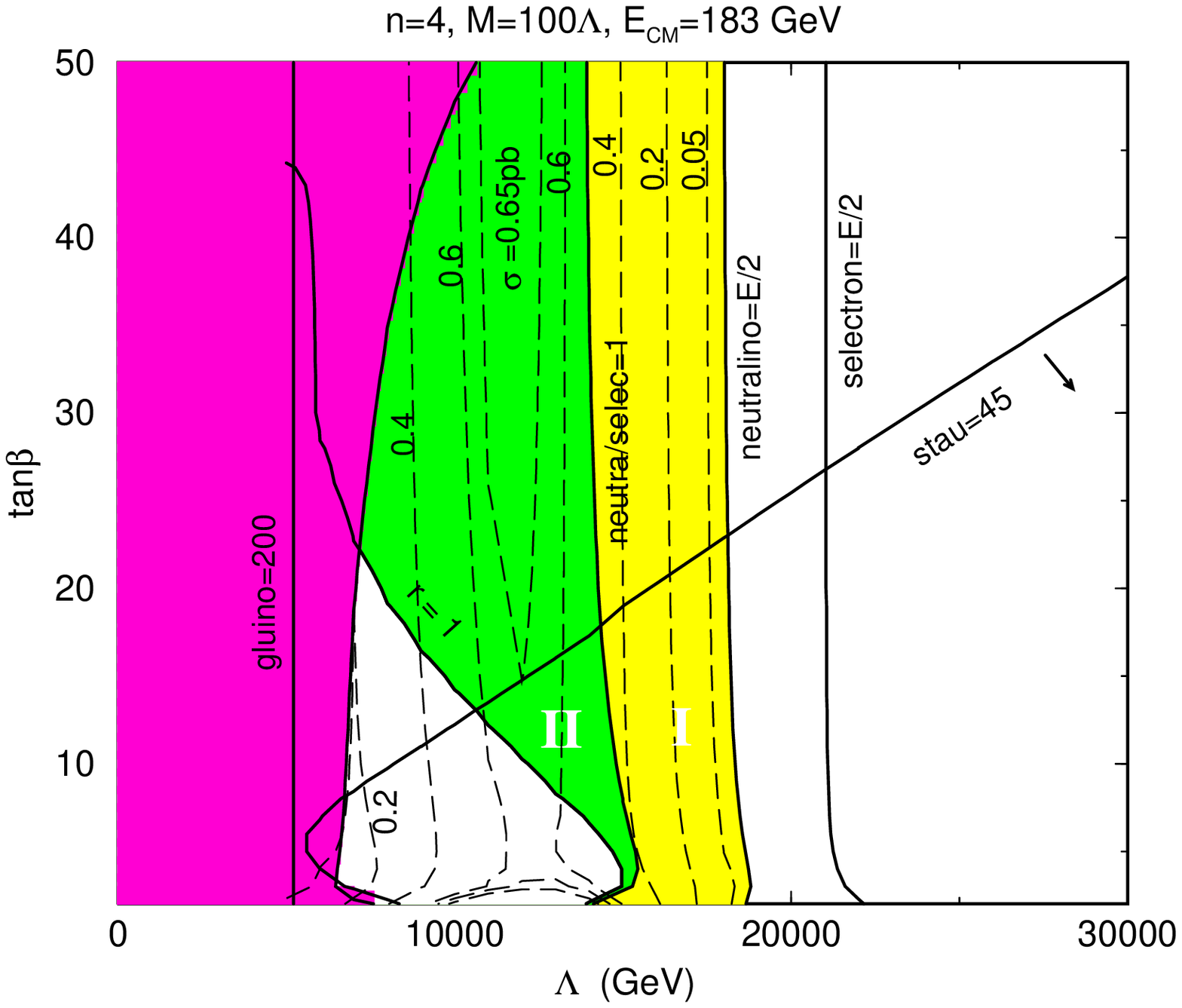}
\end{center}
\caption{}
\end{figure}

\begin{figure}[th]
\leavevmode
\begin{center}
\includegraphics[height=2.8in]{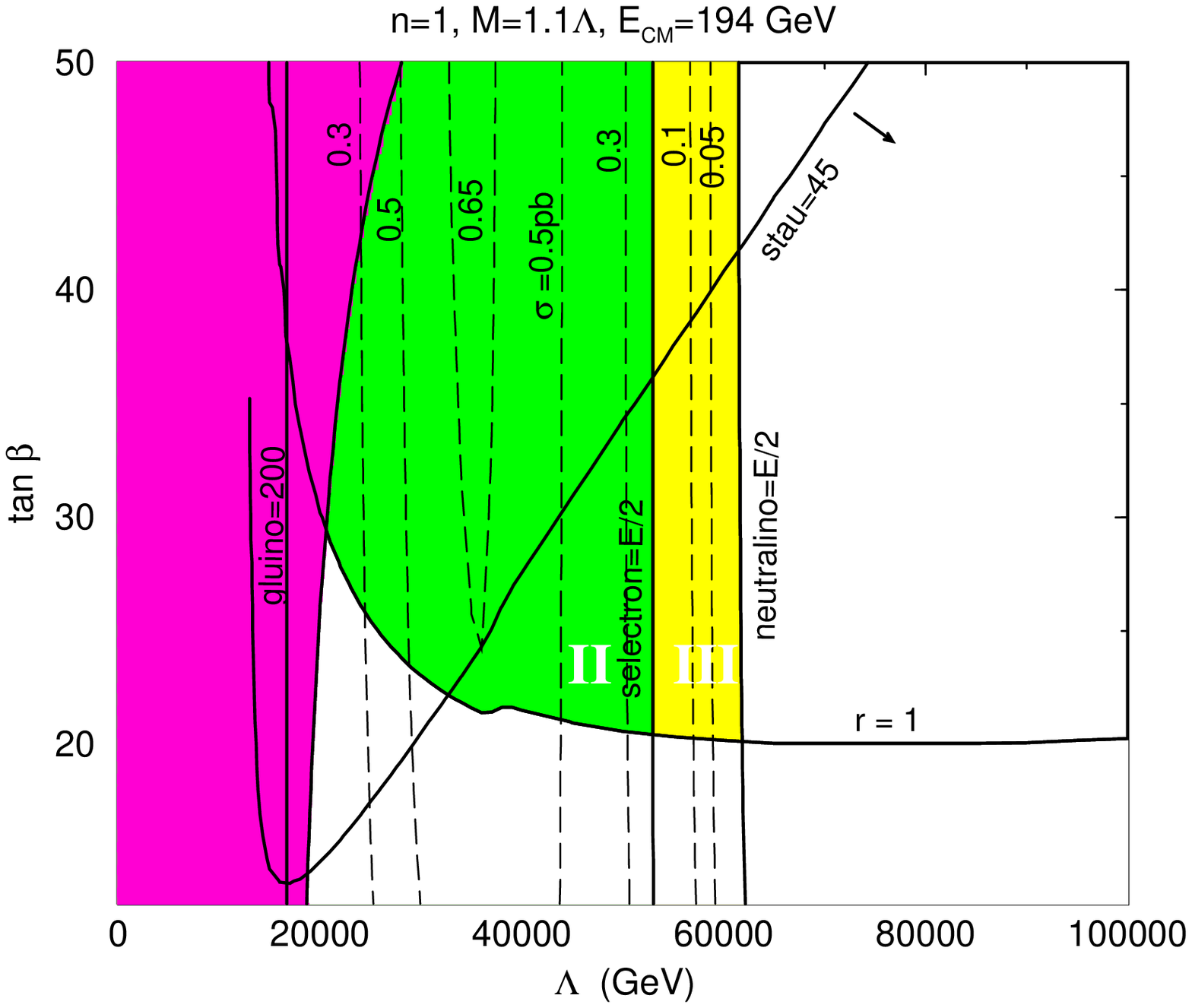}
\includegraphics[height=2.8in]{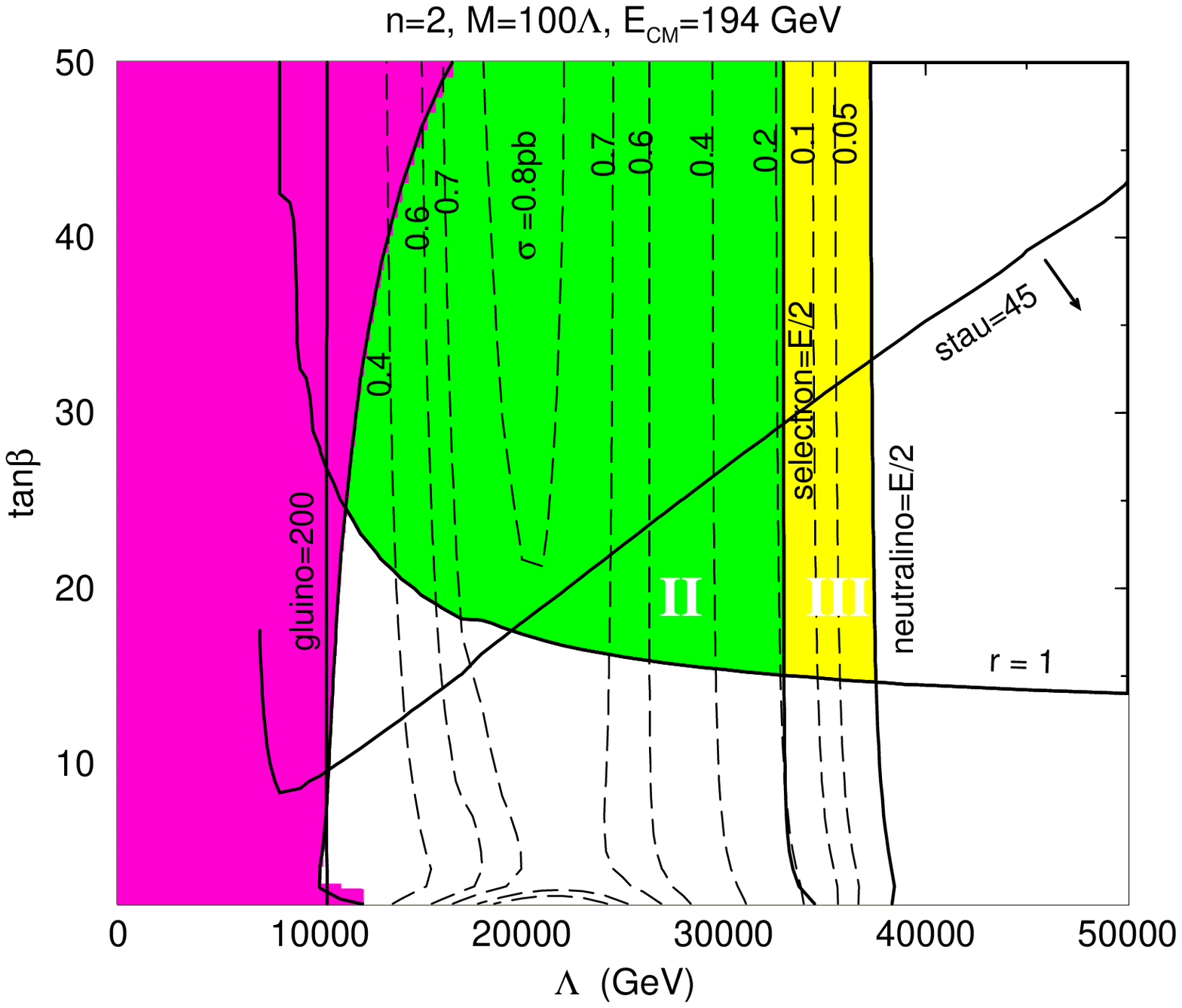}
\includegraphics[height=2.8in]{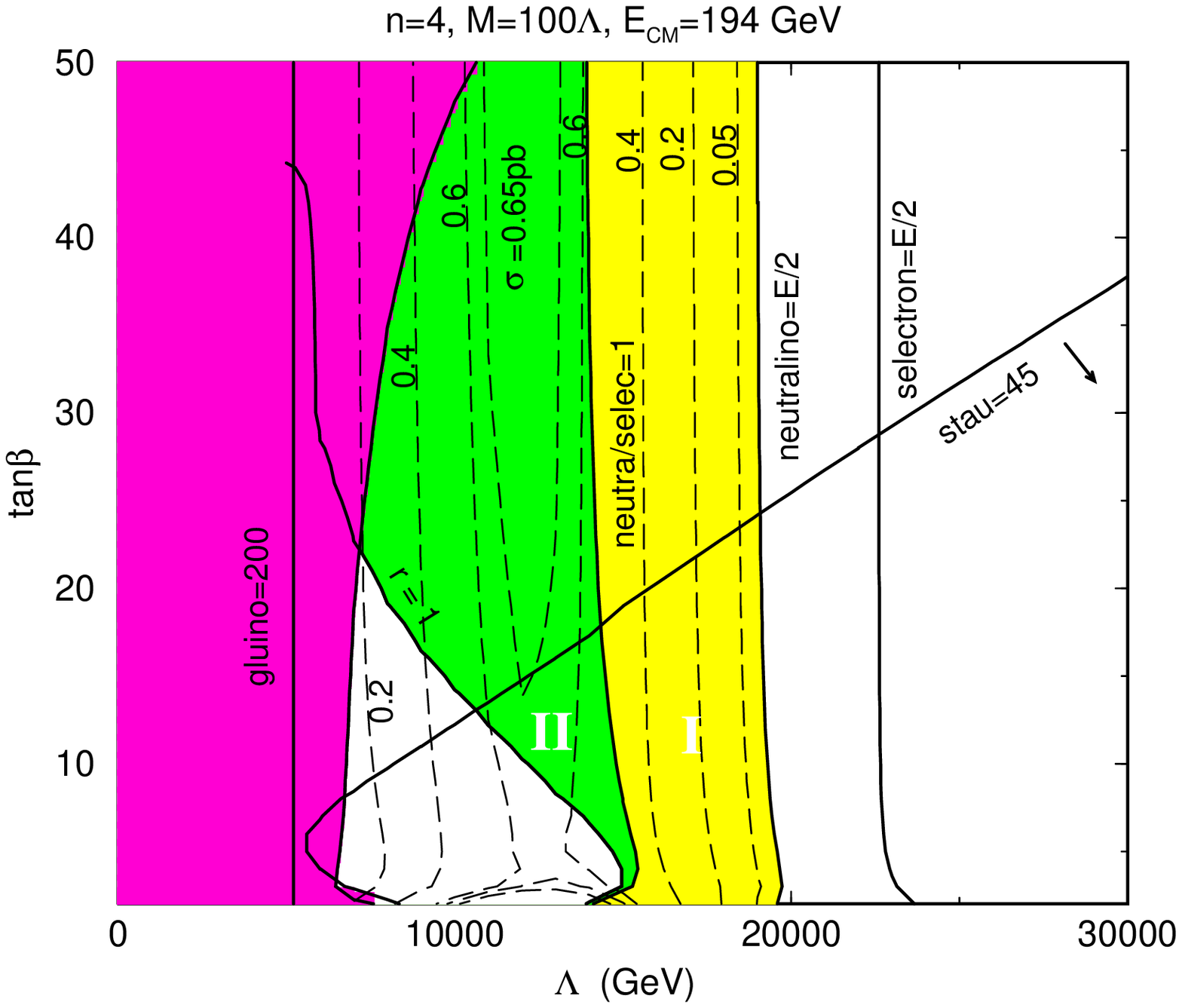}
\end{center}
\caption{}
\end{figure}

\begin{figure}[th]
\leavevmode
\begin{center}
\includegraphics[height=4.2in]{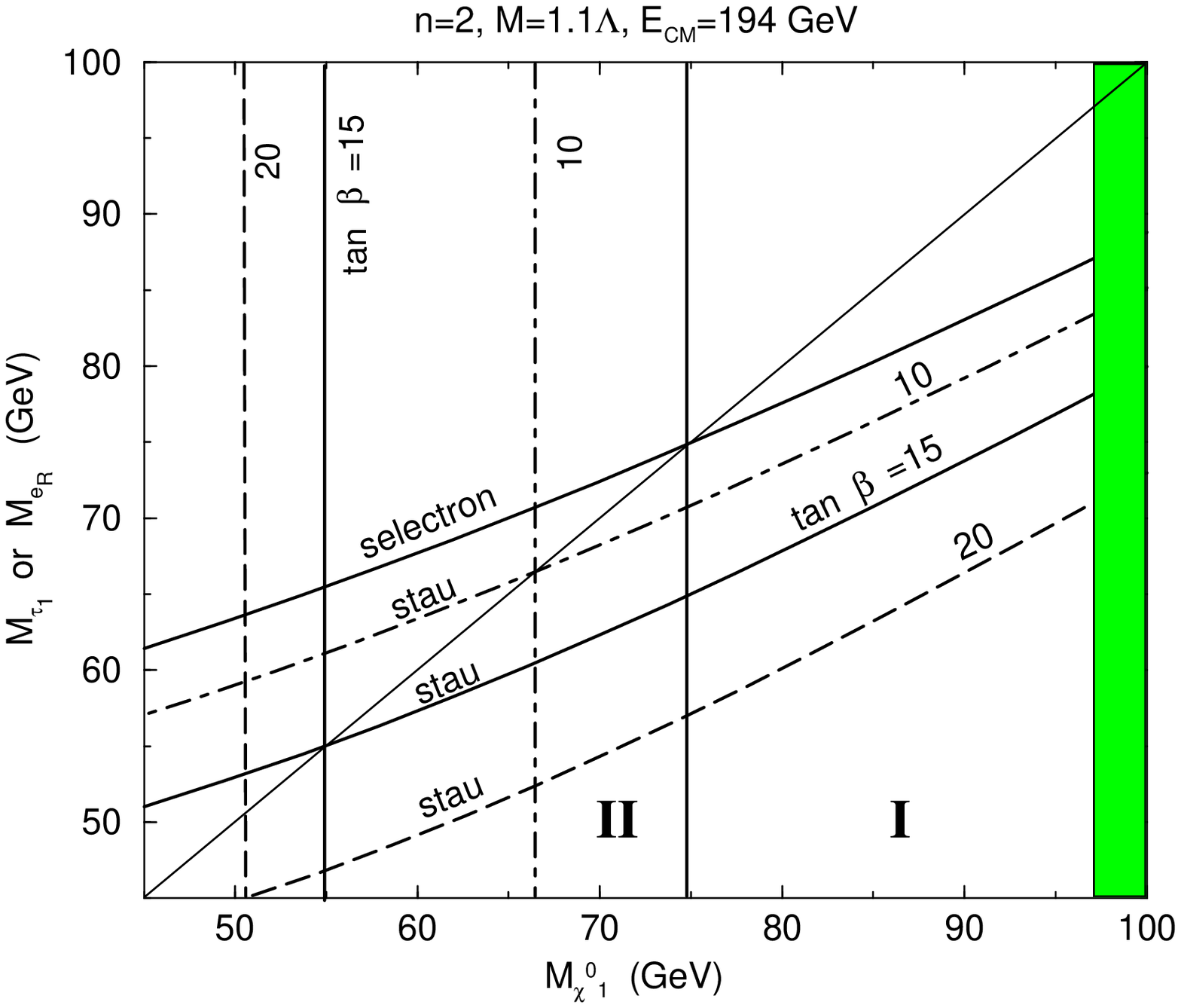}
\includegraphics[height=4.2in]{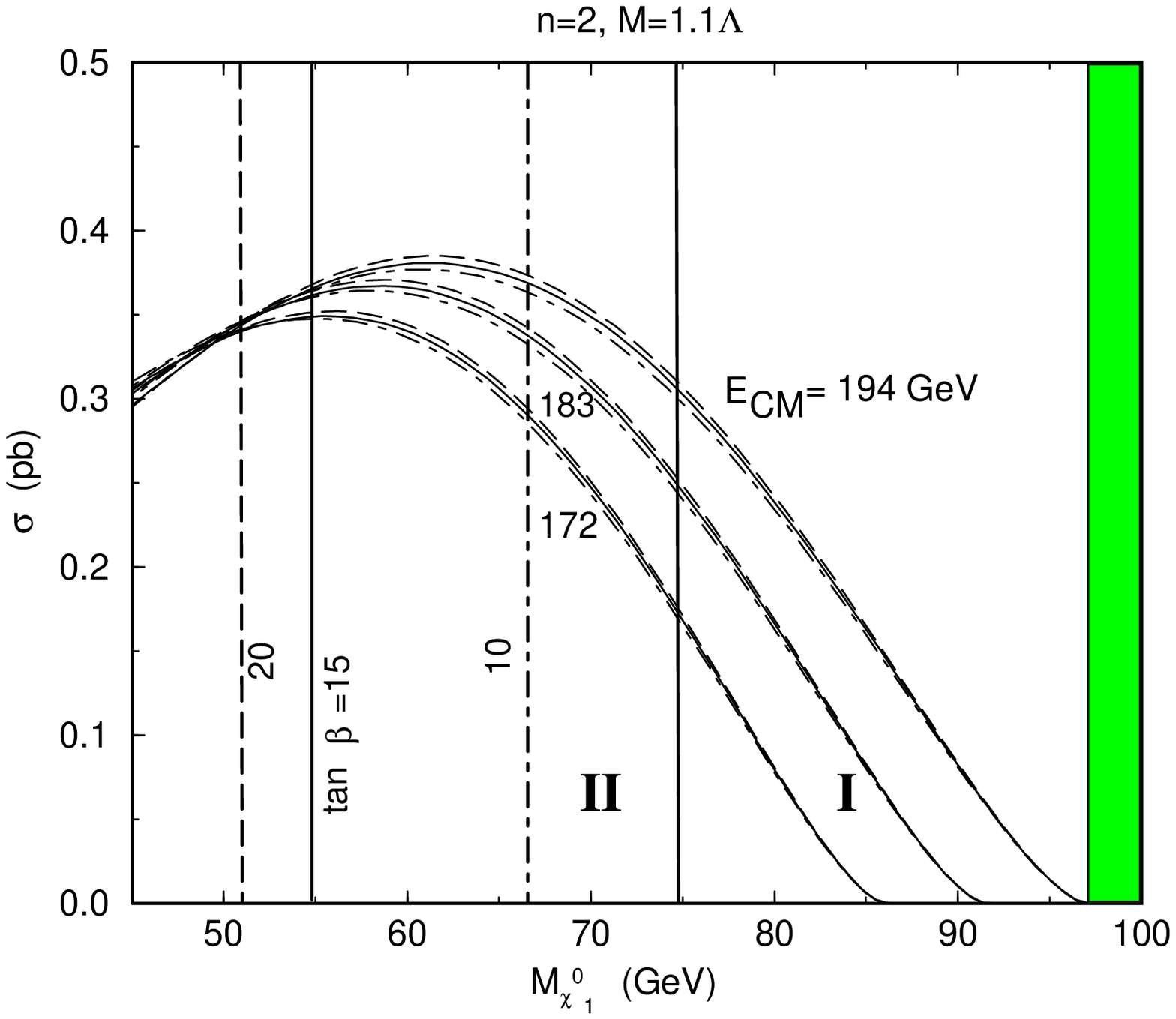}
\end{center}
\caption{}
\end{figure}

\begin{figure}[th]
\leavevmode
\begin{center}
\includegraphics[height=4.2in]{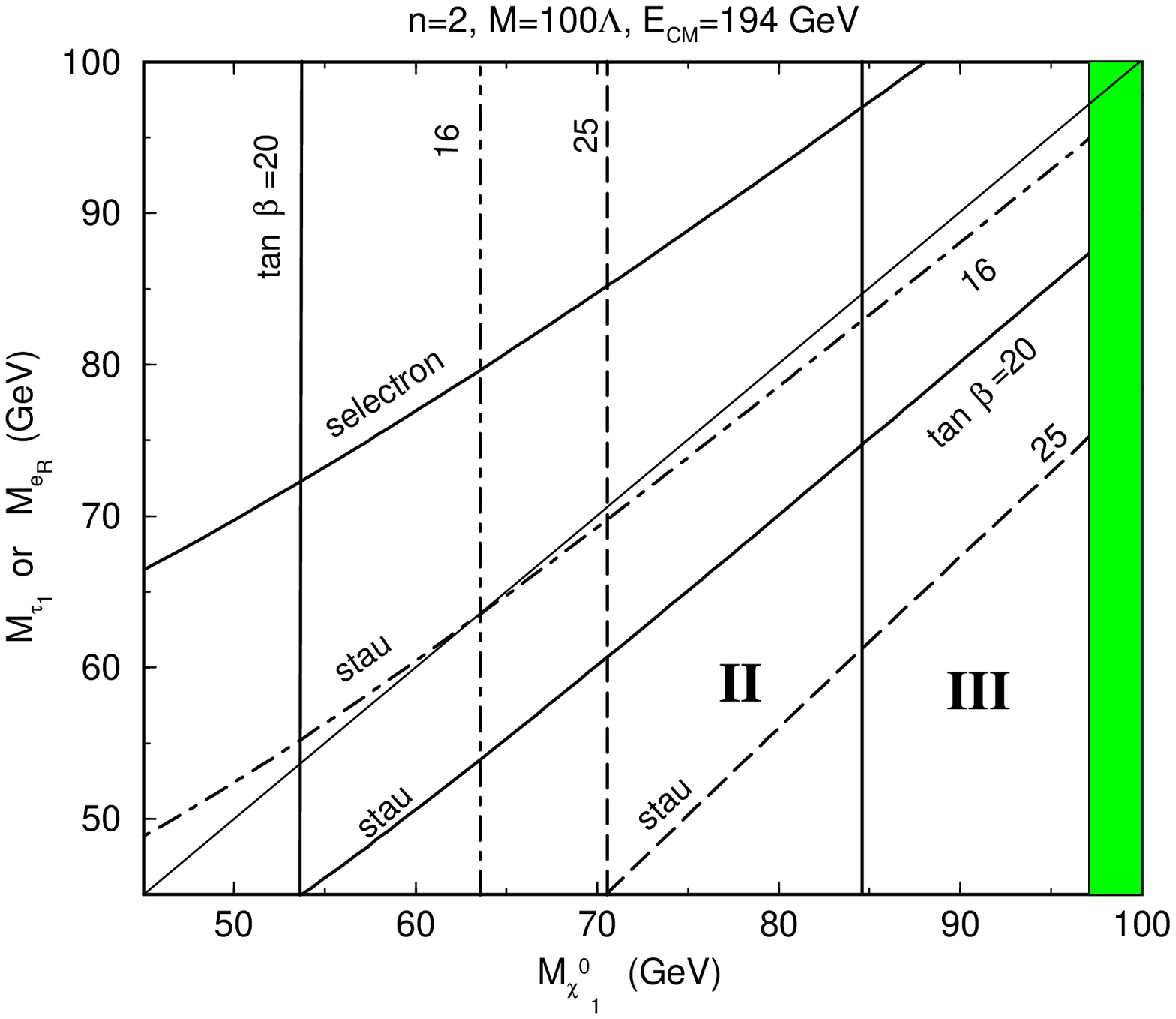}
\includegraphics[height=4.2in]{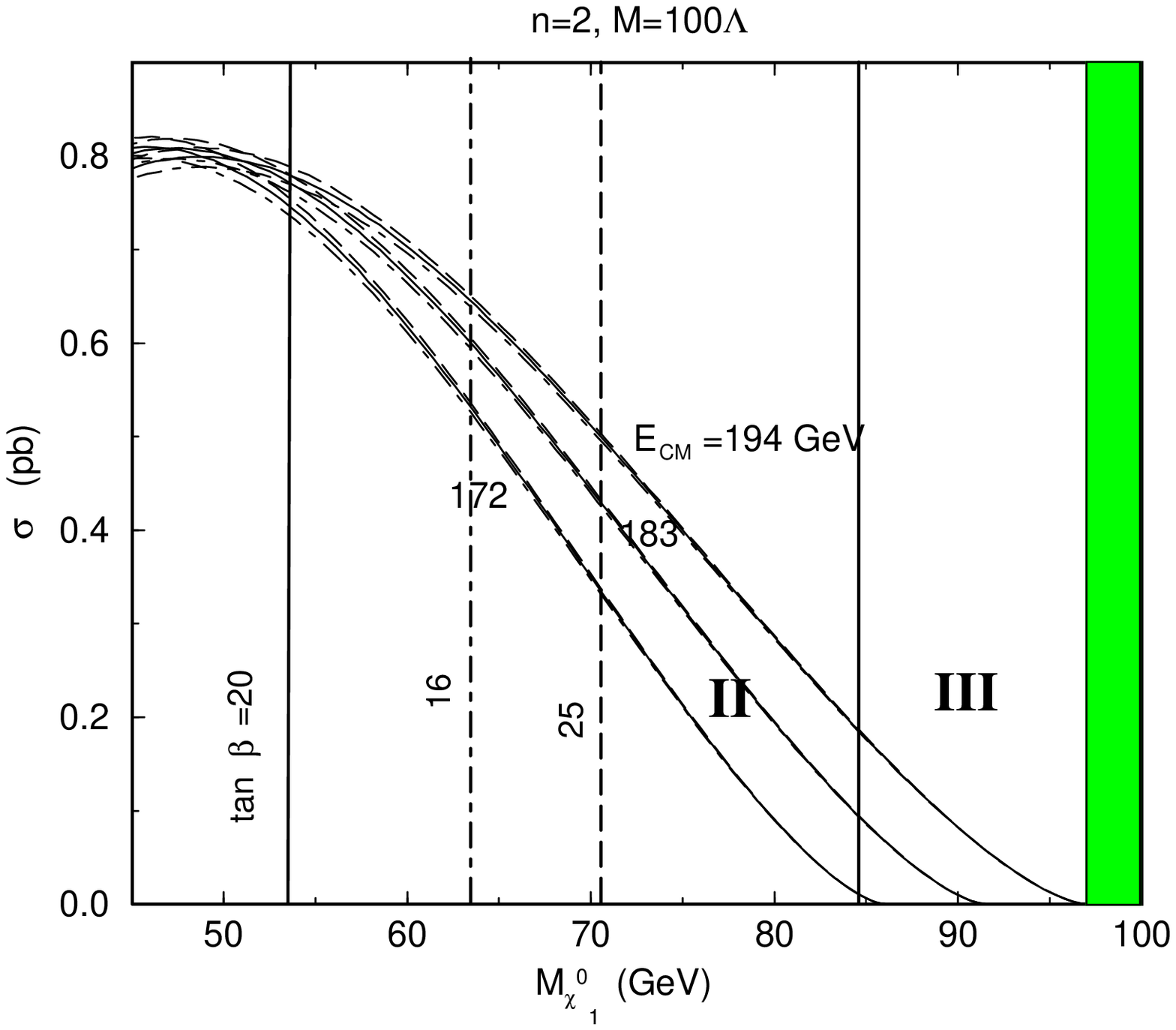}
\end{center}
\caption{}
\end{figure}


\begin{figure}[th]
\leavevmode
\begin{center}
\includegraphics[width=11cm]{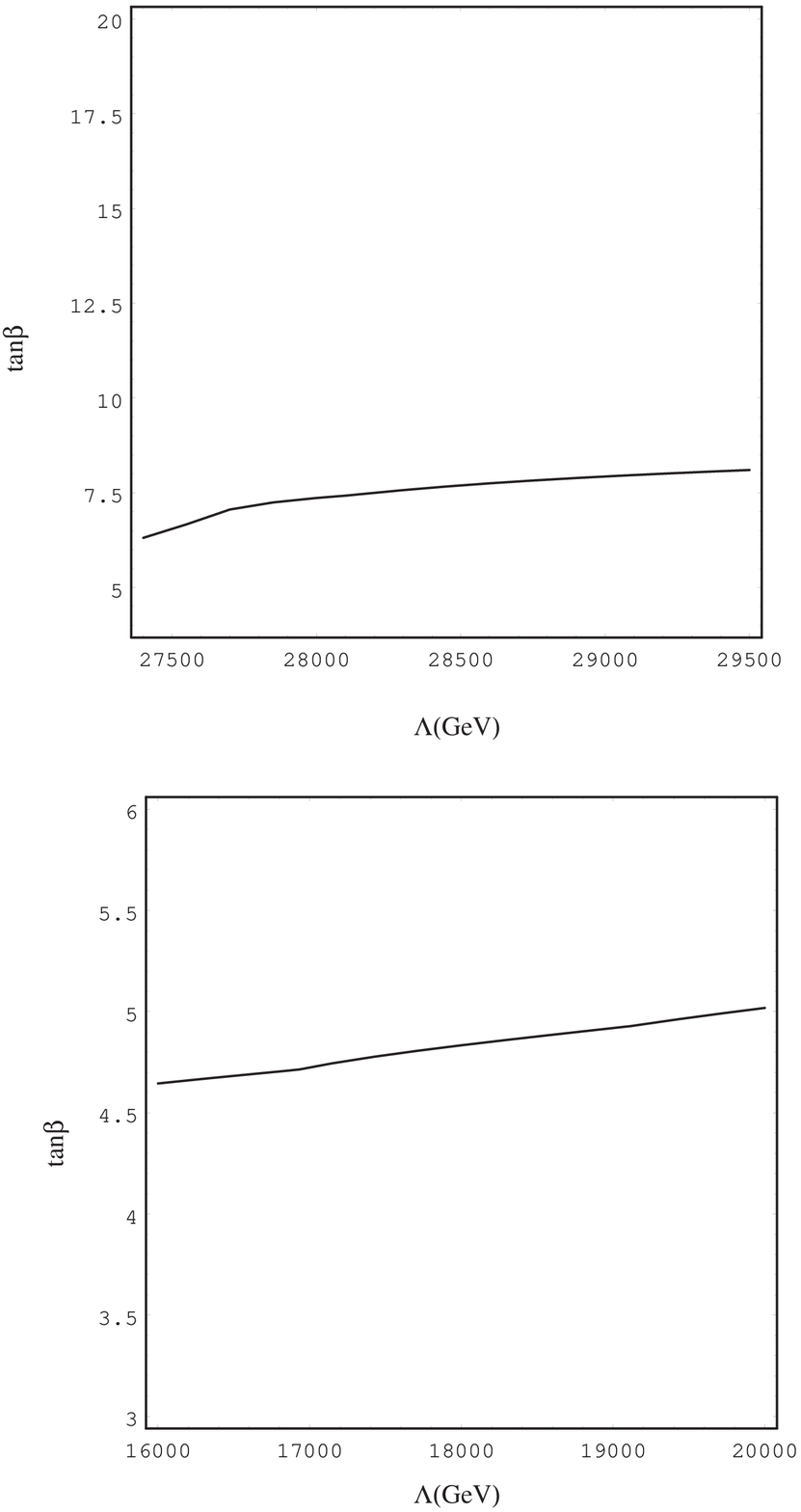}
\end{center}
\caption{}
\end{figure}

\begin{figure}[th]
\leavevmode
\begin{center}
\includegraphics[height=4.2in]{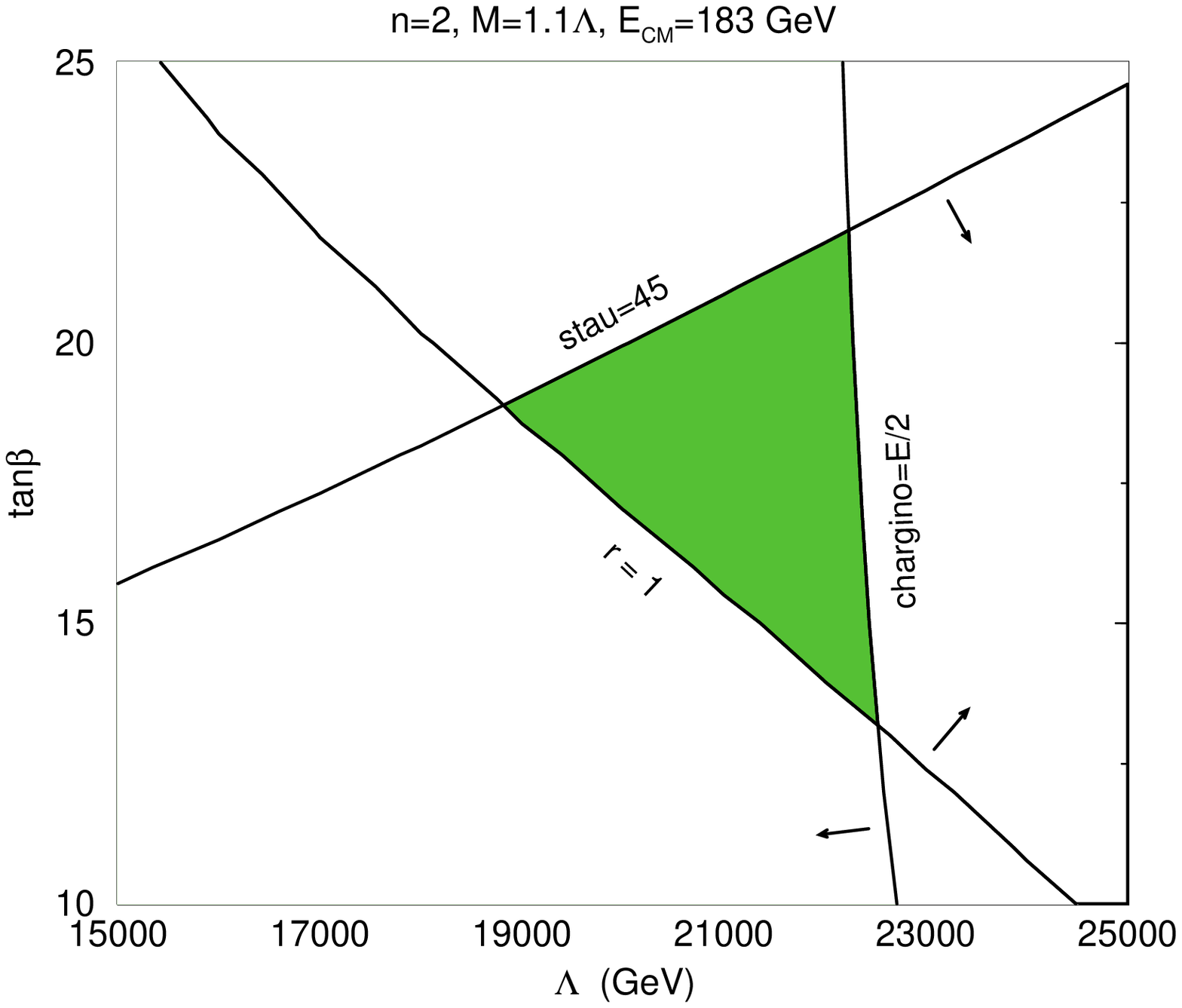}
\includegraphics[height=4.2in]{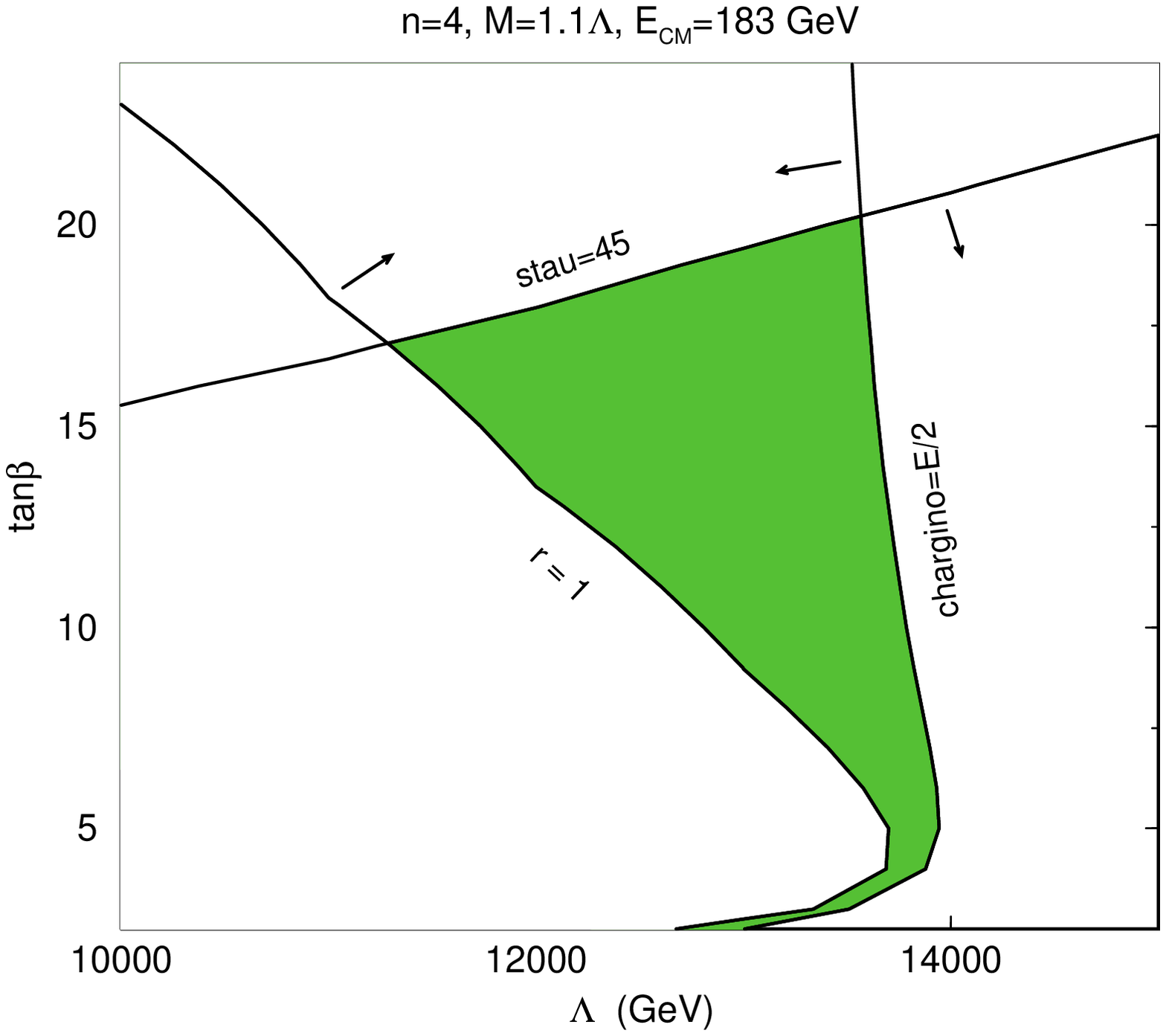}
\end{center}
\caption{}
\end{figure}


\begin{thebibliography}{[001]}
\bibitem{pol} This was first discussed by Dicus et. al in ref. \cite{we1}. See
also ref.\cite{sb} where the detailed results for the cross-sections have been
worked out.
\end{thebibliography}
\end{document}